\documentclass[namedreferences,hyperref,optionalrh,solaromanenum]{spr-sola}

\usepackage{graphicx}                    
\usepackage{color}                       
\usepackage{soul}

\chardef\us=`\_

\newcommand{\suit}{{SUIT}}
\newcommand{\js}[1]{{\color{black} #1}}
\graphicspath{{./}{pics/}}
\begin{document}
\begin{frontmatter}
\title{Test and Calibration of the Solar Ultraviolet Imaging Telescope (SUIT) on board Aditya-L1}
\author[addressref={iucaa,tezpur},corref,email={janmejoy.sarkar@iucaa.in}]{\inits{J.}\fnm{Janmejoy}~\snm{Sarkar}\orcid{0000-0002-8560-318X}}
\author[addressref={iucaa},corref,email={nived@iucaa.in}]{\inits{V.N.}\fnm{VN}~\snm{Nived}\orcid{0000-0001-6866-6608}}
\author[addressref={iucaa},corref,email={soumya@iucaa.in}]{\inits{S.}\fnm{Soumya}~\snm{Roy}\orcid{0000-0003-2215-7810}}
\author[addressref={iucaa}]{\inits{R.D.}\fnm{Rushikesh}~\snm{Deogaonkar}\orcid{0009-0000-2781-9276}}
\author[addressref={manipal},corref,email={sreejith.p@manipal.edu}]{\inits{S.}\fnm{Sreejith}~\snm{Padinhatteeri}\orcid{0000-0002-7276-4670}}
\author[addressref={uso}]{\inits{R.B.}\fnm{Raja}~\snm{Bayanna}}
\author[addressref={iucaa}]{\inits{R.K.}\fnm{Ravi}~\snm{Kesharwani}\orcid{0009-0002-2528-5738}}
\author[addressref={iucaa,cessi}]{\inits{A.R.}\fnm{A.N.}~\snm{Ramaprakash}\orcid{0000-0001-5707-4965}}
\author[addressref={iucaa,cessi},corref,email={durgesh@iucaa.in}]{\inits{Durgesh}\fnm{Durgesh}~\snm{Tripathi}\orcid{0000-0003-1689-6254}}

\author[addressref={iucaa}]{\inits{R.G.}\fnm{Rahul}~\snm{Gopalakrishnan}\orcid{0000-0002-1282-3480}}
\author[addressref={iucaa}]{\inits{B.J.}\fnm{Bhushan}~\snm{Joshi}}
\author[addressref={iucaa}]{\inits{S.S.}\fnm{Sakya}~\snm{Sinha}}
\author[addressref={iucaa}]{\inits{M.B.}\fnm{Mahesh}~\snm{Burse}}
\author[addressref={iia}]{\inits{M.V.}\fnm{Manoj}~\snm{Varma}}
\author[addressref={ursc}]{\inits{A.T.}\fnm{Anurag}~\snm{Tyagi}}
\author[addressref={ursc}]{\inits{R.Y.}\fnm{Reena}~\snm{Yadav}}
\author[addressref={iucaa}]{\inits{C.R.}\fnm{Chaitanya}~\snm{Rajarshi}}
\author[addressref={manipal}]{\inits{H.A.}\fnm{H.N.}~\snm{Adithya}\orcid{0009-0002-1177-9948}}
\author[addressref={ursc}]{\inits{A.A.}\fnm{Abhijit}~\snm{Adoni}}
\author[addressref={tezpur}]{\fnm{Gazi A.}~\snm{Ahmed}\orcid{0000-0002-0631-4831}}
\author[addressref={iia,aries}]{\inits{D.B.}\fnm{Dipankar}~\snm{Banerjee}}
\author[addressref={iucaa}]{\inits{R.B.}\fnm{Rani}~\snm{Bhandare}}
\author[addressref={leos}]{\inits{B.R.}\fnm{Bhargava Ram}~\snm{B. S.}\orcid{0000-0001-7634-1790}}
\author[addressref={iucaa}]{\inits{K.C.}\fnm{Kalpesh}~\snm{Chillal}}
\author[addressref={iucaa}]{\inits{P.C.}\fnm{Pravin}~\snm{Chordia}}
\author[addressref={iucaa,cessi}]{\inits{A.G.}\fnm{Avyarthana}~\snm{Ghosh}\orcid{0000-0002-7184-8004}}
\author[addressref={leos}]{\inits{G.G.}\fnm{Girish}~\snm{Gowda}}
\author[addressref={ursc}]{\inits{A.J.}\fnm{Anand}~\snm{Jain}}
\author[addressref={iucaa}]{\inits{M.J.}\fnm{Melvin}~\snm{James}}
\author[addressref={ursc}]{\inits{E.J.}\fnm{Evangeline~Leeja}~\snm{Justin}}
\author[addressref={iucaa}]{\inits{D.K.}\fnm{Deepak}~\snm{Kathait}}
\author[addressref={iucaa}]{\inits{A.R.K.}\fnm{Aafaque}~\snm{Khan}\orcid{0000-0002-1244-0295}}
\author[addressref={iucaa}]{\inits{P.K.}\fnm{Pravin}~\snm{Khodade}}
\author[addressref={iucaa}]{\inits{A.K.}\fnm{Abhay}~\snm{Kohok}}
\author[addressref={iucaa}]{\inits{A.K.}\fnm{Akshay}~\snm{Kulkarni}}
\author[addressref={ursc}]{\inits{G.K.}\fnm{Ghanshyam}~\snm{Kumar}} 
\author[addressref={iucaa}]{\inits{N.M.}\fnm{Nidhi}~\snm{Mehandiratta}}
\author[addressref={iucaa}]{\inits{V.M.}\fnm{Vilas}~\snm{Mestry}}
\author[addressref={iucaa}]{\inits{D.M.}\fnm{Deepa}~\snm{Modi}}
\author[addressref={ursc}]{\inits{S.M.}\fnm{Srikanth}~\snm{Motamarri}}  
\author[addressref={iia}]{\inits{N.K.}\fnm{K.}~\snm{Nagaraju}}
\author[addressref={cessi,iiserk}]{\inits{D.N.}\fnm{Dibyendu}~\snm{Nandy}}
\author[addressref={ursc}]{\inits{S.N.}\fnm{S.}~\snm{Narendra}}
\author[addressref={ursc}]{\inits{S.N.}\fnm{Sonal}~\snm{Navle}}
\author[addressref={ursc}]{\inits{N.P.}\fnm{Nashiket}~\snm{Parate}}
\author[addressref={iucaa}]{\inits{S.P.}\fnm{Sujit}~\snm{Punnadi}}
\author[addressref={ursc}]{\inits{A.R.}\fnm{A.}~\snm{Ravi}}
\author[addressref={ursc,cessi}]{\inits{K.S.}\fnm{K.}~\snm{Sankarasubramanian}}
\author[addressref={ursc}]{\inits{G.S.}\fnm{Ghulam}~\snm{Sarvar}}
\author[addressref={ursc}]{\inits{N.S.}\fnm{Nigar}~\snm{Shaji}}
\author[addressref={mps}]{\inits{S.S.}\fnm{Sami K.}~\snm{Solanki}}
\author[addressref={ursc}]{\inits{R..}\fnm{Rethika}~\snm{T}}
\author[addressref={ursc}]{\inits{K.V.}\fnm{Kaushal}~\snm{Vadodariya}}
\author[addressref={ursc}]{\inits{D.V.}\fnm{D. R.}~\snm{Veeresha}}
\author[addressref={leos}]{\inits{R.V.}\fnm{R}~\snm{Venkateswaran}}

\address[id=iucaa]{Inter-University Centre for Astronomy and Astrophysics, Post Bag 4, Ganeshkhind, Pune - 411007, Maharashtra, India}
\address[id=tezpur]{Department of Physics, Tezpur University, Napaam, Tezpur 784028, India}
\address[id=manipal]{Manipal Centre for Natural Sciences, Manipal Academy of Higher Education, Karnataka, Manipal- 576104, India}
\address[id=uso]{Udaipur Solar Observatory (USO), Udaipur, Rajasthan, India}
\address[id=cessi]{Center of Excellence in Space Sciences India, Indian Institute of Science Education and Research Kolkata, Mohanpur 741246, West Bengal, India}
\address[id=iia]{Indian Institute of Astrophysics, Koramangala, Bengaluru - 560034, Karnataka, India}
\address[id=ursc]{U R Rao Satellite Centre,  Old Airport Road Vimanapura Post, Bengaluru - 560017, Karnataka, India}
\address[id=aries]{Aryabhatta Research Institute of Observational Sciences (ARIES), Manora Peak, Nainital - 263001 Uttarakhand, India}
\address[id=leos]{Laboratory for Electro-Optics Systems (LEOS), ISRO, First Cross, First Phase, Peenya, Bengaluru- 560058, Karnataka}
\address[id=iiserk]{Department of Physical Sciences, Indian Institute of Science Education and Research Kolkata, Mohanpur 741246, West Bengal, India}
\address[id=mps]{Max Planck Institute for Solar System Research, Justus-von-Liebig-Weg 3, 37077 G\"ottingen, Germany}

\runningauthor{Sarkar et al.}
\runningtitle{Test and Calibration of SUIT}
\begin{abstract}
The Solar Ultraviolet Imaging Telescope (SUIT) on board the Aditya-L1 mission observes the Sun in the 200{--}400~nm wavelength range. This paper presents the results of various on ground and on board tests and their comparison with the specifications. Moreover, we also present the scheme for data calibration. We demonstrate that the test results are compliant with the specified figures, except the spatial resolution. Such discrepancy will limit the photometric measurements only, at a scale of 2.2" instead of 1.4" as originally envisioned. The results obtained here show that SUIT observations open up a new window for solar observations.
\end{abstract}
\end{frontmatter}

\section{Introduction} \label{s:intro}
The Solar Ultraviolet Imaging Telescope \cite[SUIT;][]{suit_main, suit_main_2} is one of the primary instruments on board Aditya-L1 mission \citep[][]{Seetha, aditya_mission}. It observes the photosphere and chromosphere of the Sun in the near and mid ultraviolet wavelength range of 200{--}400~nm, and provides full disk and partial disk images with a plate scale of 0.7"/pixel. For this purpose, it utilizes 11 science filters and a back-thinned back-illuminated 4096$\times$4096 charge-coupled device (CCD) with a 12~micron pixel size and enhanced UV sensitivity.

\begin{figure}
\centering
    \includegraphics[trim={1cm 0.5cm 1cm 5cm}, clip, width=1.0\textwidth]{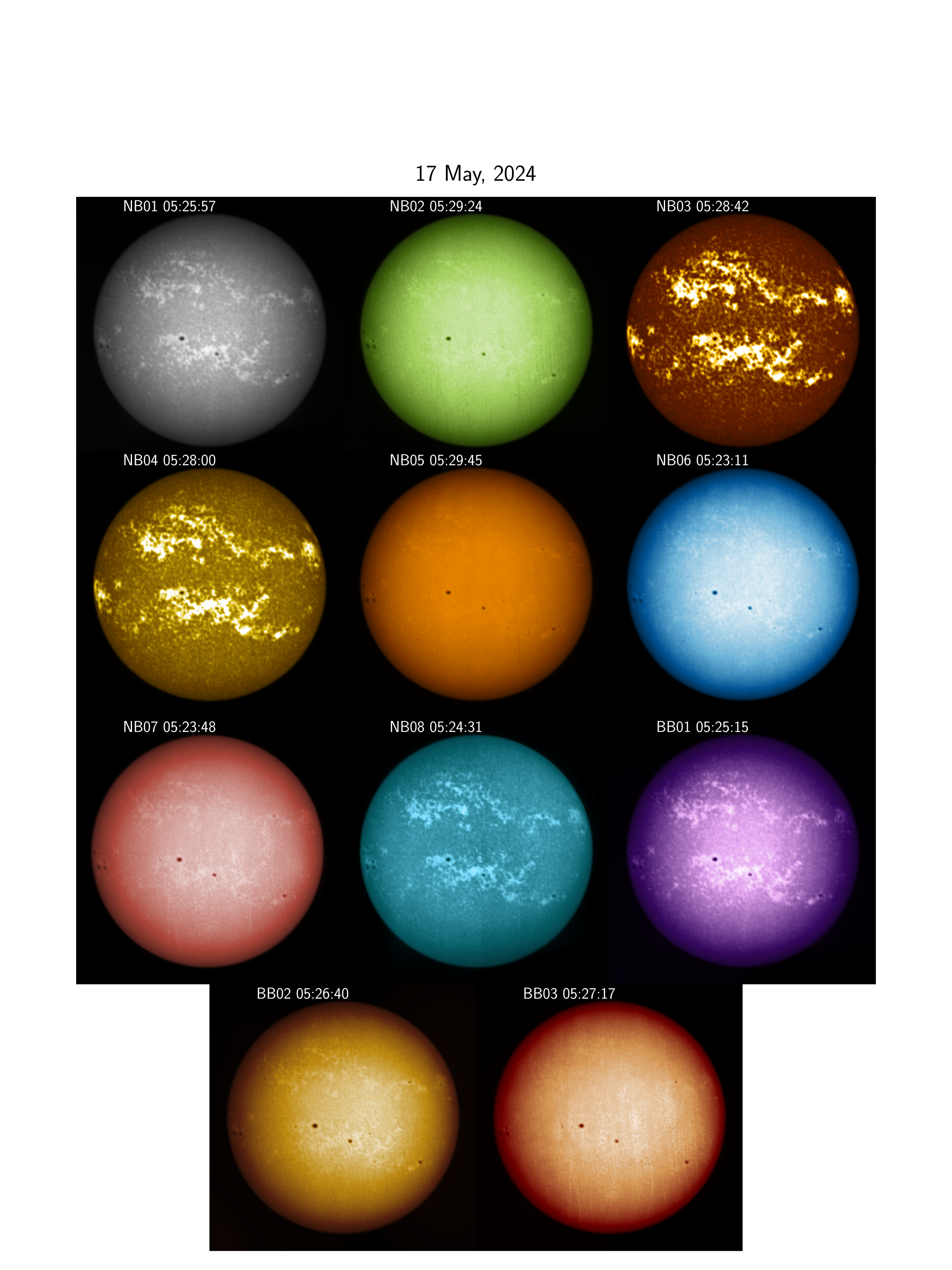}
    \caption{Full disk images in all eleven filters recorded by SUIT on May 17th, 2024.}\label{full_disk}
\end{figure}

The spacecraft was launched on Sep 02, 2023 and was subsequently put into a halo orbit around L1 point on Jan 06, 2024. The telescope was switched on on Nov 20, 2023, with the first light taken on Dec 06, 2023. Since then, the telescope has been recording observations for various tests, calibration, and verification of the telescope. Fig.~\ref{full_disk} displays fully calibrated science-ready full disk images recorded on May 17th, 2024, in all eleven filters as labeled.

In this paper, we describe both on-ground and in-orbit tests and calibrations performed on SUIT. The rest of the paper is structured as follows. In \S\ref{s:design} we provide a brief overview to the instrument. We discuss the pre-flight tests performed on ground and the results in \S\ref{s:gtest}. The in-orbit test and results are discussed in \S\ref{s:onboard}. In \S\ref{s:calibration}, we discuss the calibration procedures for generating the science-ready Level 1 data products from raw Level 0 data.

\section{Instrument Overview}\label{s:design}
\begin{figure}
    \centering
    \includegraphics[width=0.9\linewidth]{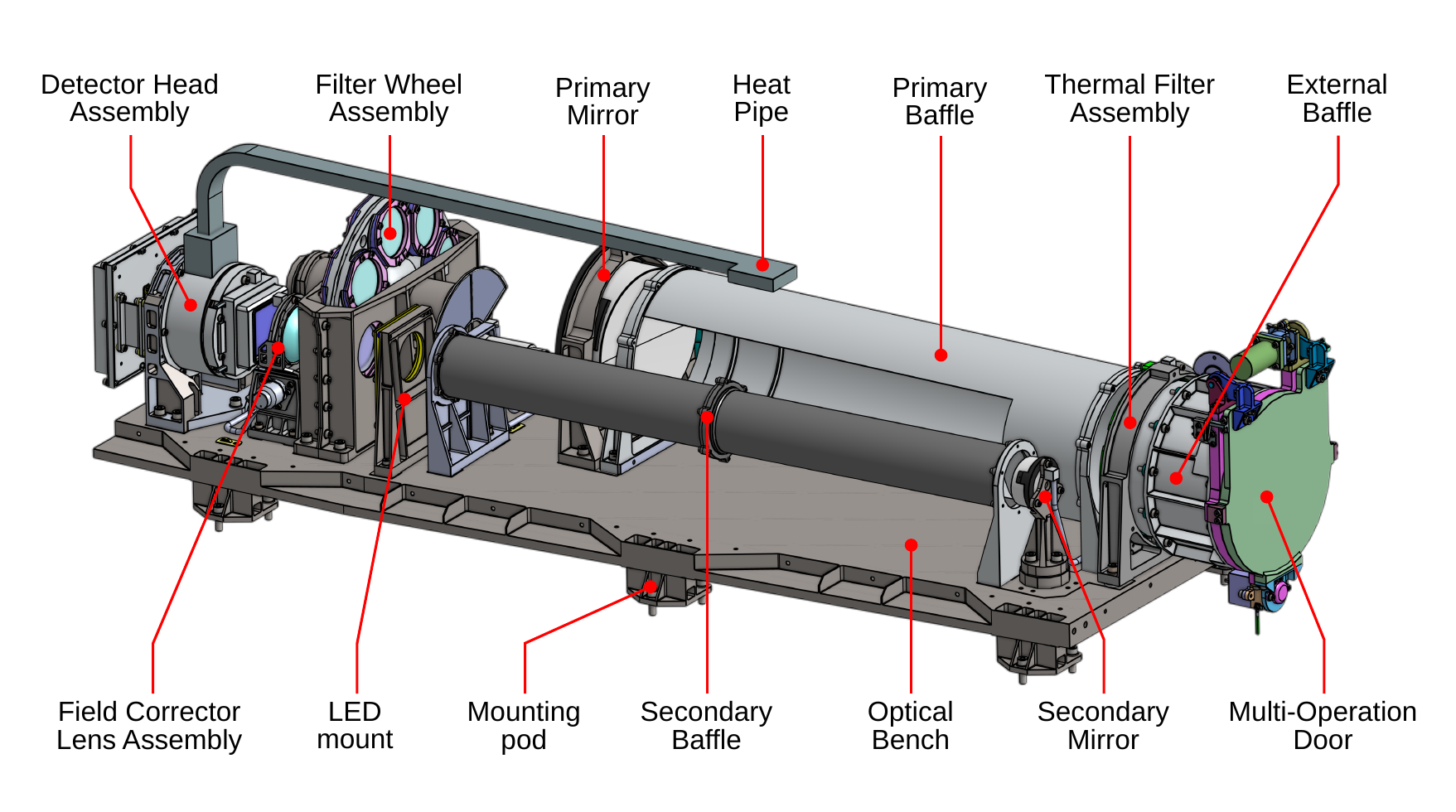}
    \caption{Schematic diagram of the Solar Ultraviolet Imaging Telescope. The internal components are labeled.}
    \label{fig:suit_drawing}
\end{figure}
\begin{table*}[!ht]
\caption{SUIT Instrument Characteristics}\label{tab:instrument}      
\begin{tabular}{l|r}
\hline
\textbf{Parameter} 	& \textbf{Value}   \\
\hline
Telescope Design    & f/24.8 off-axis Ritchey Chr\'{e}tien \\
Wavelength range    & 200 nm - 400 nm.\\
Bandpass            & 8 Narrow-band, 3 Broad-band \\
Entrance Aperture 	& 146 mm \\
Detector            & 4096$\times$4096, back-thinned, back-\\ 
                    & illuminated, UV-enhanced CCD\\
Plate Scale			& 0.7"/pixel @ 12 $\mu$m pixel\\
Exposure times 		& 0.1 to 1.4 s  \\
\js{Fastest Cadence}& \js{6 sec in RoI mode}\\
\hline
\end{tabular}
\end{table*}

\suit\ is an off-axis Ritchey-Chr\'etien type solar telescope operating in the 200{--}400~nm wavelength band. The instrument is mounted on the top deck of the Aditya-L1 spacecraft \citep[see][]{Seetha,aditya_mission}. Fig.~\ref{fig:suit_drawing} shows the schematic diagram of the telescope with its internal components. The critical parameters of the telescope are given in Table~\ref{tab:instrument}. The front aperture of the optical cavity has a multi-operational door, which is operated during the spacecraft momentum dump events to prevent any contamination. The entrance aperture filter, called the thermal filter, is designed to transmit 0.01\% of the incident sunlight into the optical cavity \citep{thermal, thermal2}. This helps to regulate the temperature of the optical cavity and prevents saturation of the CCD.

\begin{table}
\begin{tabular}{l c c c r}
\hline
\textbf{Science}  &	\textbf{Central} & \js{\textbf{Bandpass}} &\textbf{Science}\\
\textbf{Filter}	 &	\textbf{Wavelength (nm)} & \js{\textbf{(nm)}} &\textbf{target}\\
\hline
NB01    & 214.0 	& \js{11.0} & Continuum\\
NB02 	& 276.6		& \js{0.4} 	& Mg~\rm{II}~k blue wing \\
NB03 	& 279.6 	& \js{0.4} 	& Mg~\rm{II}~k\\
NB04 	& 280.3		& \js{0.4} 	& Mg~\rm{II}~h\\
NB05	& 283.2		& \js{0.4} 	& Mg~\rm{II}~h red wing\\
NB06 	& 300.0 	& \js{1.0} 	& Continuum\\
NB07 	& 388.0		& \js{1.0} 	& CN Band\\
NB08	& 396.85 	& \js{0.1} 	& Ca~\rm{II}~h\\
BB01 	& 220.0		& \js{40.0}	& Herzberg Continuum \\
BB02 	& 277.0 	& \js{58.0} & Hartley Band\\
BB03 	& 340.0		& \js{40.0} & Huggins Band\\
\hline
\end{tabular}
\caption{Science filters, central wavelength\js{,} and science target of the 11 science filters for SUIT. \js{Refer to \cite{Sarkar2024} for further details on filters and their characterization.}} \label{sc_comb_fil} 
\end{table}

The telescope has an entrance aperture of 146~mm, with primary and secondary mirrors having diameters of 141~mm and 49~mm, respectively. Sunlight is reflected off the primary and secondary mirrors and is incident on the science filters mounted on two stacked filter wheels with 8 slots each. The two filter wheels generate eight narrow and three broad bandpasses by stacking specific filters (see Table~\ref{sc_comb_fil}). The filtered light passes through a field correction lens before getting imaged on the CCD. \js{Detailed information on the filters and their characterization can be found in \cite{Sarkar2024}}. The field correction lens is mounted on a piezo-electric linear stage, allowing $\pm$~3~mm movement along the optical axis.

For calibration purposes, \suit\ is equipped with an array of eight 355~nm and 255~nm LEDs each. This array of 16 LEDs is mounted before the science filters in the optical path. Light from the LEDs pass through the filters and field corrector lens before falling on the CCD. For more details see \cite{suit_main_2,suit_main}. The telescope also comprises of three baffles, {\it viz.} external, primary and secondary, as labeled in Fig.~\ref{fig:suit_drawing}. 

\section{Ground Tests and Results}\label{s:gtest}

\subsection{Field of View (FOV)}\label{sec:gnd_fov}

In order to check for any obstruction due to baffles in light passing through the \suit\ optics, we performed a test to measure the field of view (FOV). This test was performed before mounting the CCD on the focal plane. For this purpose, we used a Fizeau interferometer and a f/3.5 transmission sphere with the beam focused on the focal plane of \suit. The light passes through the telescope optics and is retro-reflected by a plane mirror kept at the entrance aperture of the telescope. The tilt of this mirror is varied to check the vignetting for various field angles. In the lab, the test was performed for $\pm$~23~mm, corresponding to $\pm$~0.37$^\circ$, from the center of the imaging plane, in horizontal and vertical directions. Our analysis of this test reveals that there was no obstruction throughout this range. Note that while the original requirement for the FOV is 0.39$^\circ$, due to mechanical / optical limitations, we could not measure beyond 0.37$^\circ$ during ground tests.

\subsection{Measurement of Point Spread Function and Encircled Energy} \label{sec:lab-psf}
The ground test setup for the telescope was established in a clean room at the ISRO Satellite Integration and Testing Establishment (ISITE) in Bengaluru, India. We loaded the telescope in a vacuum chamber and cooled the CCD, maintaining a temperature of $-55^\circ C$ using liquid nitrogen micro-dosing through a cooling jacket, which was mounted on the CCD heat pipe. 

For this test, we also needed a collimated source. For this purpose, we have used the flight-spare model of the telescope as a collimator. The light source is kept at the collimator focal plane to get a collimated beam. This collimator was placed outside the vacuum chamber and was aligned with the telescope such that the collimated light entered the telescope through the vacuum chamber viewport.
    
Given the specification of the instrument design, the diffraction-limited diameter of the point spread function (PSF) is $\sim \mathrm{12~\mu m}$ at a wavelength of 200~nm. A target smaller than half the diffraction limit of the telescope would appear as a point source to the telescope optics. Therefore, a pinhole of $\mathrm{5~\mu m}$ diameter was placed at the focal plane of the collimator and illuminated by a Xenon Arc Lamp. The light from this point source gets collimated and passes through the entire \suit\ optics to be finally imaged on the {\suit} CCD. We measured the PSF in all eleven wavelength bands by rotating the two filter wheels.

\begin{figure}
\centering
    \includegraphics[trim={0.5cm 3cm 1cm 3cm}, clip, width=0.9\linewidth]{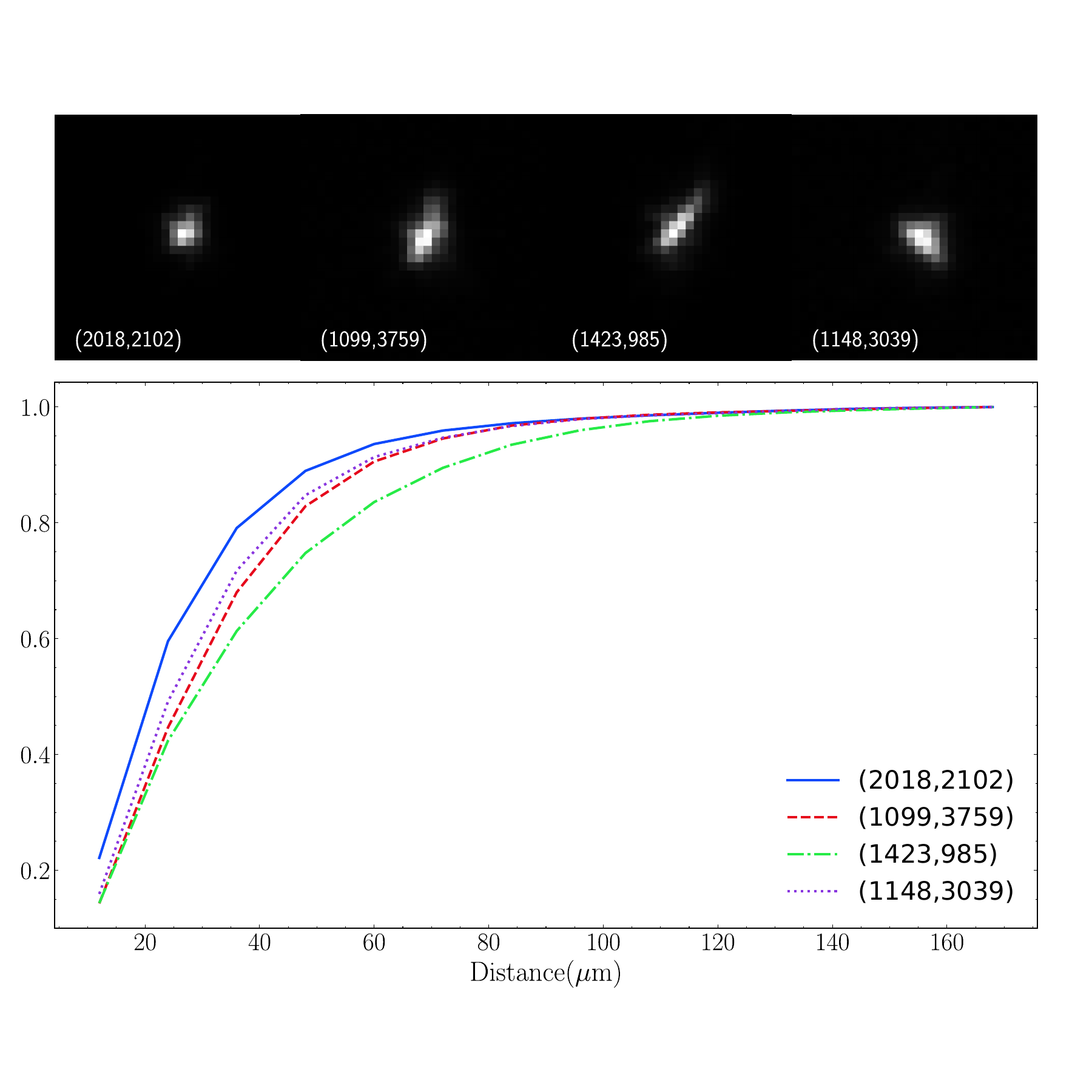}
    \caption{\js{Encircled energy plot for {\suit} PSF measured in the lab at various locations of the CCD for the NB07 filter. The marked (x,y) coordinates denote the position of the PSF on the CCD. These measurements are representative of the PSF in all observation bands.}}
    \label{fig:ensquared_e}
\end{figure}

A reliable metric for the sharpness of the PSF is to obtain the number of pixels enclosing 80\% of the total energy of the PSF. A smaller radius signifies a tight PSF, which indicates sharper images. Here we present the PSF in the NB07 filter, centered at a wavelength of 388 nm. To obtain the 80\% encircled energy in a PSF spot, it is mandatory to characterize and remove the background level from the measurements. Here, we estimate the background by taking the median of counts recorded within a 5-pixel annulus, more than 100 pixels away from the PSF spot in all directions. Fig.~\ref{fig:ensquared_e} top panels show the measured PSF at various positions on the CCD. The centroid of the PSF is noted in each panel of the figure. The bottom panel shows the encircled energy as a function of radial distance from the centroid for the PSFs at different positions, as marked. The encircled energy is calculated by adding the background subtracted counts in 15-pixel annuli centered around the centroid using the \texttt{aperture\_photometry} function publicly available in the \texttt{photutils} package \citep{photutils}. The central PSF (blue solid line) rises most sharply, while the PSF patch farthest away from the CCD center (green dot-dashed line) rises slowest, indicating a poorer PSF in the latter case. The plot reveals that near the CCD center, 80\% encircled energy is obtained at the radius of $\sim 35~\mu m$, corresponding to $\sim$ 2.04~arcsec.

It is important to note that the collimator used in this experiment is the flight-spare model of the telescope, which has similar optical properties as the flight model. Therefore, the obtained 80\% encircled energy radius is due to the cumulative effect of the aberrations from the telescope and the collimator. The wavefront errors of both telescopes were tested interferometrically and showed similar performance. Given their almost identical nature, the aberrations must be added in quadrature. Therefore, dividing the 80\% encircled energy radius by $\sqrt{2}$ gives the contribution from just the telescope. This gives us a radius of $35/\sqrt{2}~=~24.75~\mu m$, which is $\sim 2$ pixels on the {\suit} CCD. \js{As {\suit} primarily uses reflective optics, the measurements in one wavelength band can be considered to be representative of the PSF in all observation bands.}

We further note that the PSF reported here is the worst case scenario given that the wavefront errors introduced by the viewport are not considered. 

\subsection{Plate Scale and Modulation Transfer Function (MTF)} \label{sec:lab-platescale}
Plate scale refers to the angular coverage of one pixel in the imaging system. This is validated by observing an object of known angular size and measuring the number of accommodating pixels. For this purpose, we have used a USAF resolution target and used the same experimental setup as described in \S\ref{sec:lab-psf}, except for the pinhole being replaced by a USAF target at the collimator focus and a modified arrangement to illuminate the target evenly. The USAF target has vertical and horizontal lines of various thicknesses and spacings. Our analysis used the 16 lines per millimeter pattern on the USAF target, where each line is equivalent to 3.65" in angular size.

\begin{figure}
        \centering
        \includegraphics[width=0.7\linewidth]{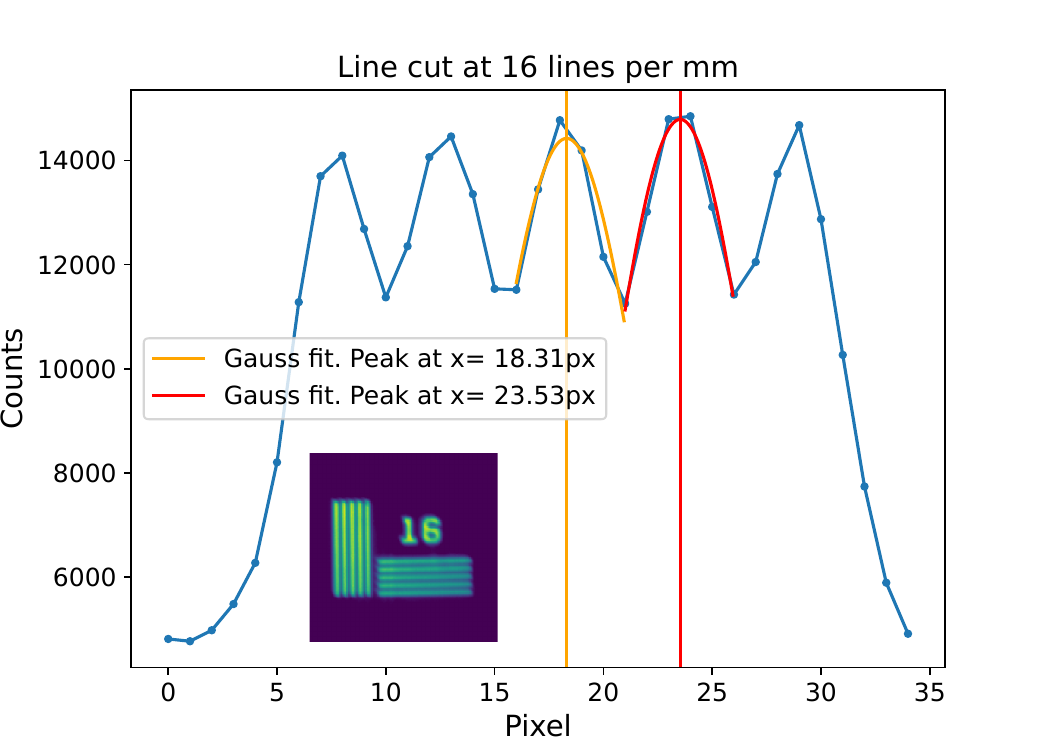}
        \caption{Line profile of 16 lines per mm pattern on USAF resolution target. The image of the target is shown here as an inset.}
        \label{fig:plate_scale}
    \end{figure}

Fig.~\ref{fig:plate_scale} plots the intensity cut through the USAF target showing various line profiles. To obtain the distance between two peaks, we have fitted Gaussians to two consecutive peaks and obtained their center. The difference between the center of these Gaussians is the separation between two lines, which is then compared with the actual separation between these lines. From this measurement, we find that the plate scale is 0.698"/px, which is almost identical to the design specification (0.7"/px) as given in Table~\ref{tab:instrument}.

The MTF of an optical system represents the contrast as a function of spatial scale/frequency. Here, the contrast is measured with a varying number of lines per mm in a USAF resolution target using the optical setup mentioned above. The percentage of contrast is measured as,
\begin{equation} \label{eqn:contrast}
    C= \frac{B-D}{B+D} \times 100\%
\end{equation}
where,
\begin{itemize}
    \item \( C \) denotes the contrast percentage, which quantifies the difference in intensity between the brighter and darker regions in the USAF target images.
    \item \( B \) represents the brightness or intensity of the bright regions, through which light is passing.
    \item \( D \) signifies the brightness or intensity of the darker region, or the opaque region of the line pattern.
\end{itemize}

\begin{figure}
        \centering
        \includegraphics[width=0.7\linewidth]{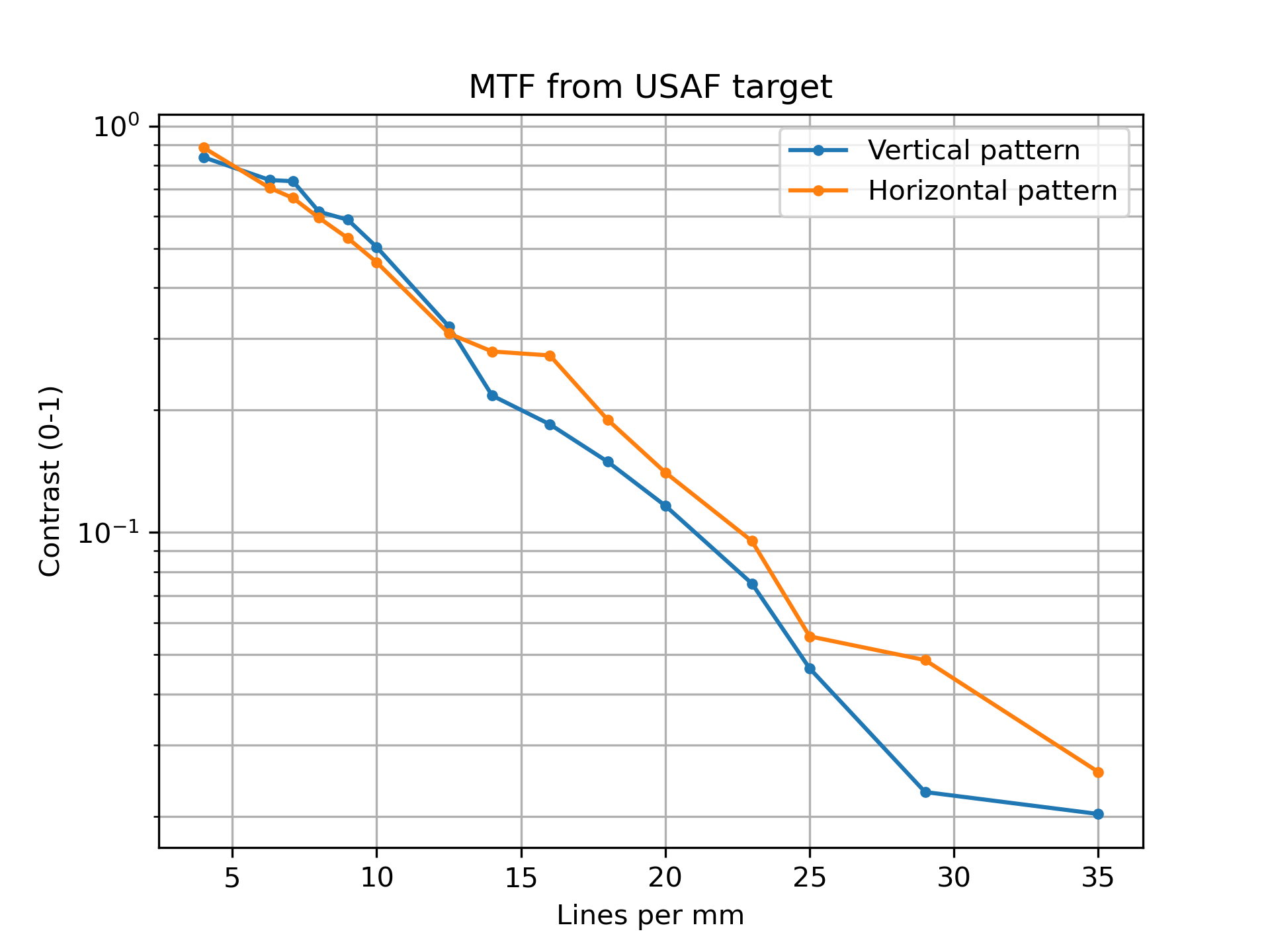}
        \caption{Modulation Transfer Function (MTF) curve for vertical and horizontal patterns on the USAF target.}
        \label{fig:mtf}
\end{figure}

Fig.~\ref{fig:mtf} plots the measured contrast as function of lines per mm, both for vertical (blue curve) and horizontal (orange curve) patterns. The plot shows that the telescope delivers 10\% contrast for spatial structures in both vertical and horizontal directions at 23 lines per mm USAF target. This corresponds to 2.5", a larger angular distance than originally anticipated. Note that the original requirement for the telescope is to deliver 10\% contrast at a 1.4" spatial scale. However, as noted earlier, this measurement includes the aberrations due to the telescope and the collimator, therefore, representing the upper limit for resolution.

\subsection{Read noise and Bias}
Read noise refers to the amount of signal the CCD electronics generates while reading and writing an image to the memory. The Level~0 images have 64 columns on either side of the image, added as overscan pixels. These overscan pixels are generated by \suit\ electronics and are not affected by light falling on the CCD. 

The fluctuation over the set bias value in the overscan pixels provides the read noise level. We used these overscan pixels during the thermo-vacuum tests and measured the read noise as 8~e$^{-}$, which is better than the required specification of 10 $e^-$. Note that the CCD bias was set to $\sim 2500$ counts during ground measurements.

\subsection{Photometric Calibration}
For performing the photometric calibration of \suit, we have simulated the instrument throughput for all 11 bandpasses using a composite Sun-as-a-star spectrum from SOLSTICE \citep{mcclintock05} and SOLSPEC \citep{thuillier09} with the transmission/reflectance of all its optical elements along with the quantum efficiency and gain of the detector and allied electronics. We compare these values with those obtained from experimental measurements as described below. 
    
Note that the experimental setup is similar to that described in \S\ref{sec:lab-psf}, except that for this test, we use a monochromator to feed light of a specific wavelength and bandpass at the focal plane of the collimator with an optical fiber. The central wavelength of the monochromatic light is chosen based on the bandpass to be calibrated. We use a National Institute of Standards and Technology (NIST) traceable photo-diode to measure the absolute power of the light emerging from the collimator to get absolute power measurements. The light from the collimator is fed into the SUIT optical cavity and imaged. We compare these results with those obtained by throughput modeling. The ratios of the experimental results are within 20\% of the simulated results as given in Table~\ref{tab:throughput}. This is except for NB07, which can be attributed to the low spectral step size of the solar spectrum used for modeling the instrument throughput. A detailed description of this test can be found in \cite{jj_photo}.

\begin{table} 
\begin{tabular}{c|c}
\hline
\textbf{Science filter} & \textbf{Ratio of $\frac{DN_{measured}}{DN_{simulated}}$} \\
\hline
NB02 & 1.08 \\
NB03 & 1.19 \\
NB04 & 0.95 \\
NB05 & 0.94 \\
NB06 & 0.98 \\
NB07 & 1.91 \\
NB08 & 1.10 \\
BB03 & 1.18 \\
\hline
\end{tabular}
\caption{Comparison of the data numbers derived from the throughput model (using SOLSTICE and SOLSPEC data) with the data numbers inferred from the measurements.} \label{tab:throughput}
\end{table}

\section{On-Board Test, Observations and Results}\label{s:onboard}

\subsection{Plate Scale}

Plate scale refers to the angular coverage of one pixel in the imaging system. This is validated by observing a subject of known angular size and measuring the number of accommodating pixels. 

For this purpose we use a full disk image taken using the Ca~\rm{II}~h (396.85~nm) filter. 
We measure the diameter of the solar disk in this image using our limb-fitting algorithm and compare it with the published values of the solar diameter obtained using Ca~\rm{II}~K images recorded from the ground (\cite{2018A&A...616A..64M}). Here we assume that the formation heights of the Ca~\rm{II}~H and Ca~\rm{II}~K lines are approximately the same. 
We also account for the change in the angular size of the Sun for Ca~\rm{II}~K observations made from the Earth and Ca~\rm{II}~H observations from the Lagrange-1 point. From our observations, we compare the angular size of the Sun in Ca~\rm{II}~K with the diameter observed by {\suit} in pixels. The plate scale is measured to be 0.69"/px, which is identical to designed plate scale (0.7"/px).

Note that {\suit} is an off-axis telescope. Therefore, the optics show distortion at high field angles. This causes the plate scale to vary at high field angles from the field center. As a result, significant elongation of features is seen as we move from the center to the corners of the CCD. The change is noticeable for field angles $>0.3^\circ$. For example, we find from optical simulation that the plate scale is $\sim 0.54"/px$, when distortion correction is not applied at the extreme corner of the telescope field of view ($ 0.39 ^\circ \times 0.39 ^\circ$). 

\begin{figure}
\centering
\includegraphics[trim={0cm 0.5cm 0.2cm 0cm}, clip, width=0.95\linewidth]{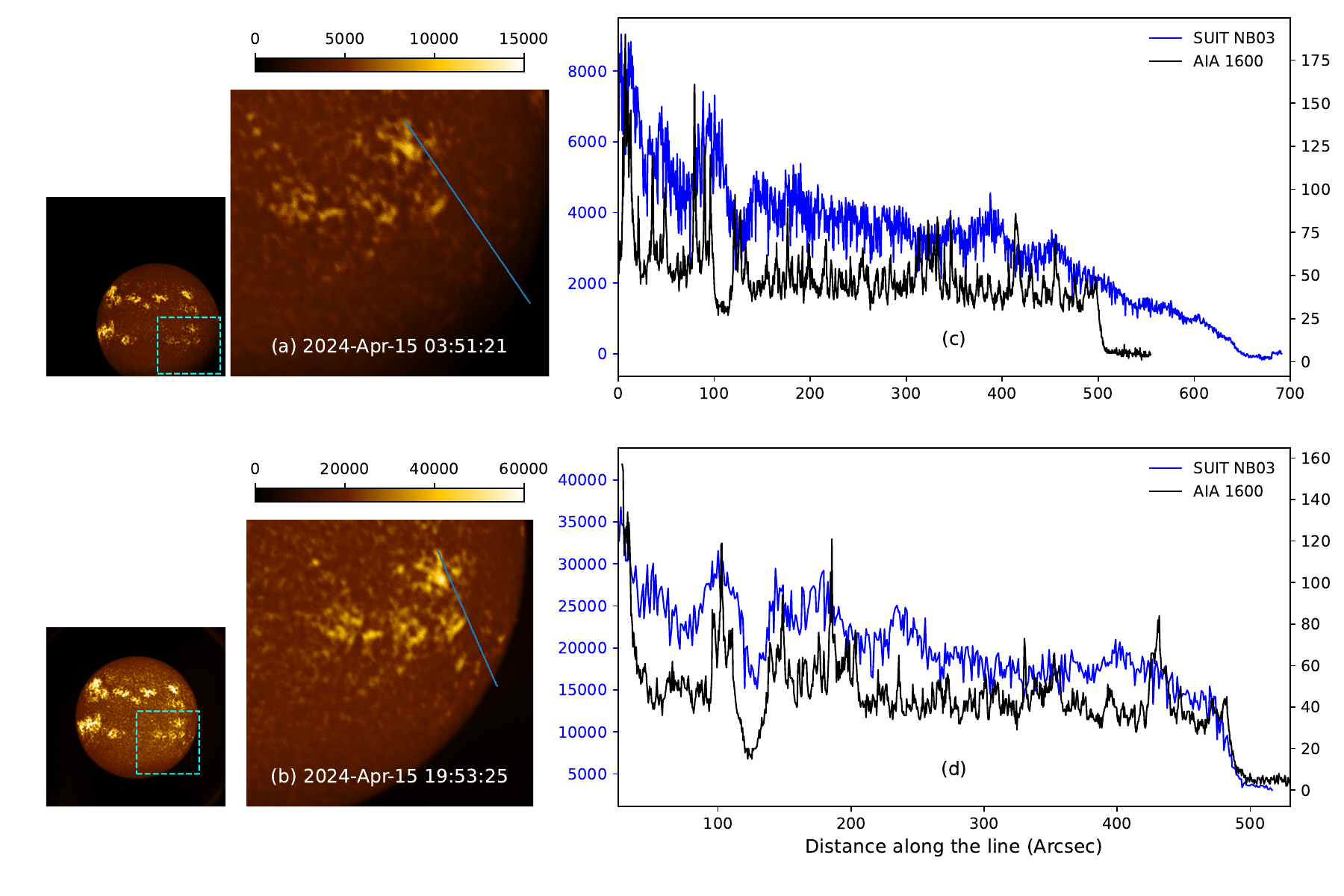}
\caption{SUIT NB3 images captured on the same day roughly \js{16 hours} apart. \js{Frames a and c: solar image offset to the lower right corner of the CCD, frames b and d: solar image centered on the CCD.} The intensity profile across the blue line in panels a and b are plotted in panels c and d, respectively, along with the corresponding AIA 1600 intensity. We notice that the brightness decays rapidly at the solar limb in panel d, but varies gradually in panel c, due to vignette beyond $0.39^\circ$ from the frame center.}
\label{fig:corner-dist}
\end{figure}

\subsection{Field of View} \label{sec:ob_fov}
    \begin{figure}
        \centering
        \includegraphics[width=0.5\linewidth]{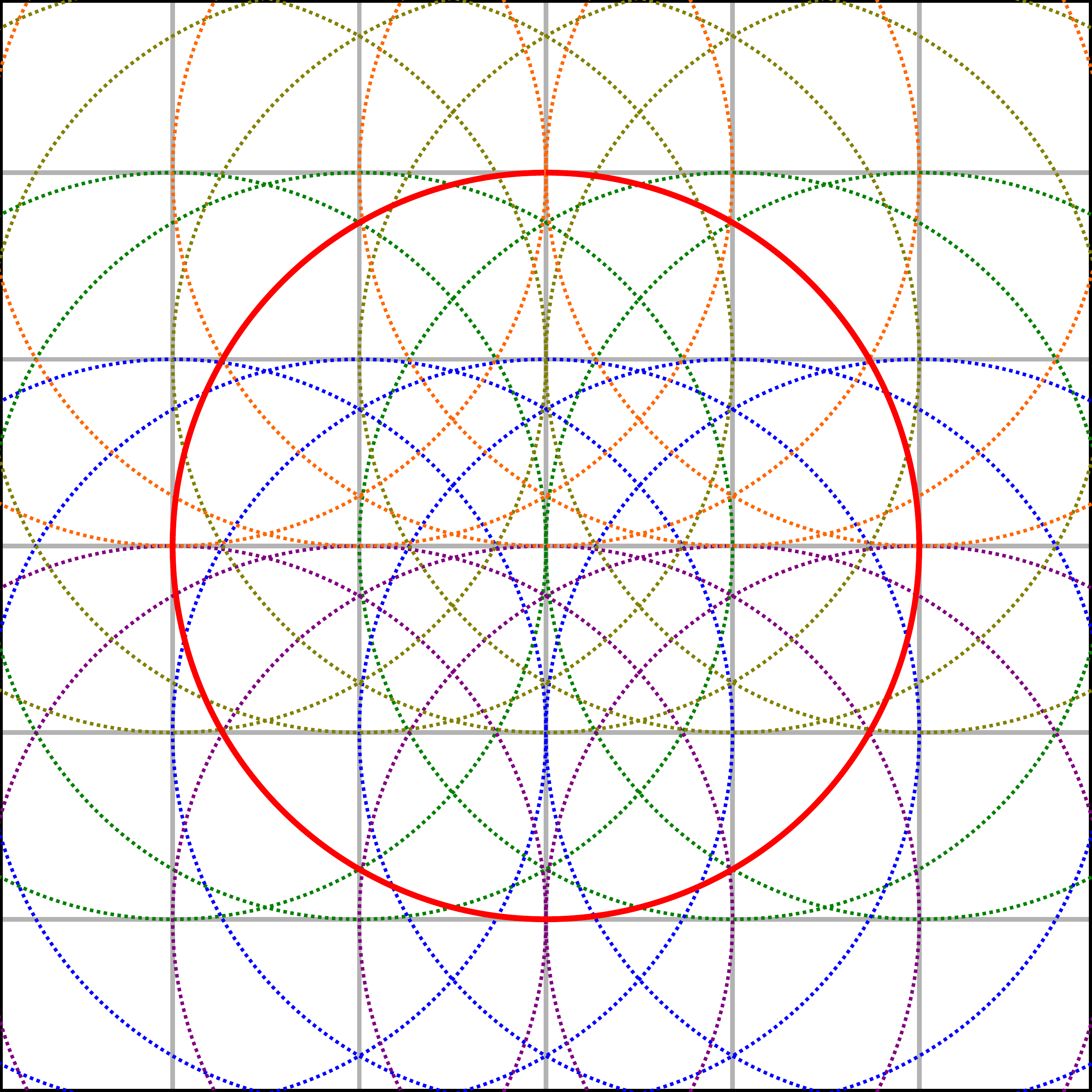}
        \caption{Solar off pointing scheme. Images are taken at offsets of $\pm 8$ and $\pm 16$ arcminutes in vertical and horizontal directions about the field center. Each circle represents the solar disk for each pointing of \suit.}
        \label{fig:off-pointing}
    \end{figure}
\begin{figure}
        \centering
        \includegraphics[width=0.7\linewidth]{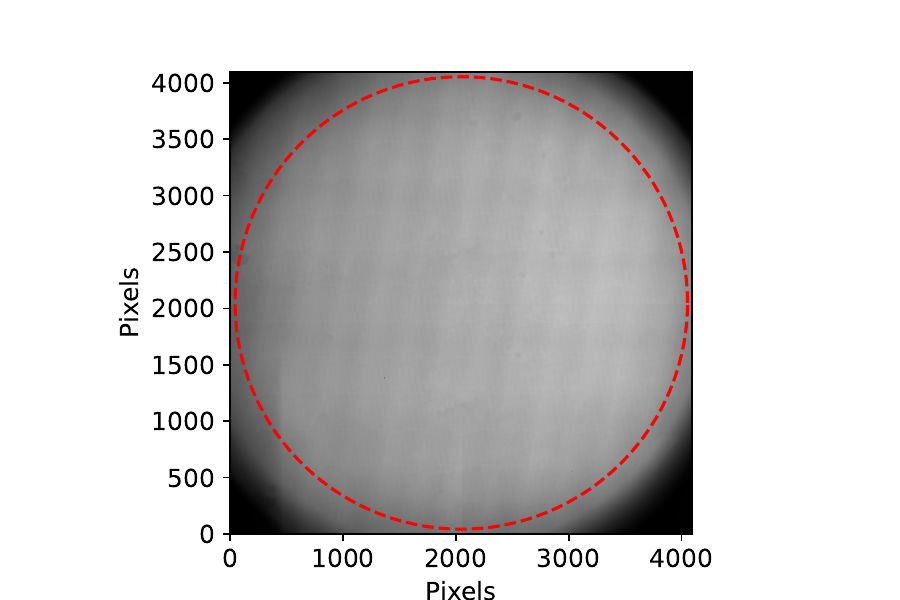}
        \caption{Off-pointed Sun images taken by \suit\ are merged using a maximization method. The red dashed line shows a circle of radius $0.39^\circ$, the field of view of \suit. It is noticeable that there is no vignetting within the circle, denoting the \suit\ field of view is unobstructed.}
        \label{fig:vignette_profile}
    \end{figure}

The FOV validation is performed after measuring the \suit\ plate scale. The satellite is off-pointed in pitch and roll axes in steps of 8 arcmins according to the scheme illustrated in Fig.~\ref{fig:off-pointing}. Therefore, the Sun is imaged in various regions of the CCD to map the non-uniformity of illumination and the nature of vignetting. We combine these images using a maximization algorithm, where the maximum value of each pixel in a stack of images is written to a blank grid of pixels. This evenly spreads the intensity of the solar disk across the image plane, giving us flat illumination across the field of view. Using this image we can determine the vignette profile of {\suit} and measure the unobstructed field of view (see Fig.~\ref{fig:vignette_profile}). The red dashed circle corresponds to $0.39^\circ$ radius from the center of the frame. It is clear from the image that the {\suit} field of view is unobstructed within this circle, which corresponds to the designed FOV.

\begin{figure}
    \centering
    \includegraphics[width=1\linewidth]{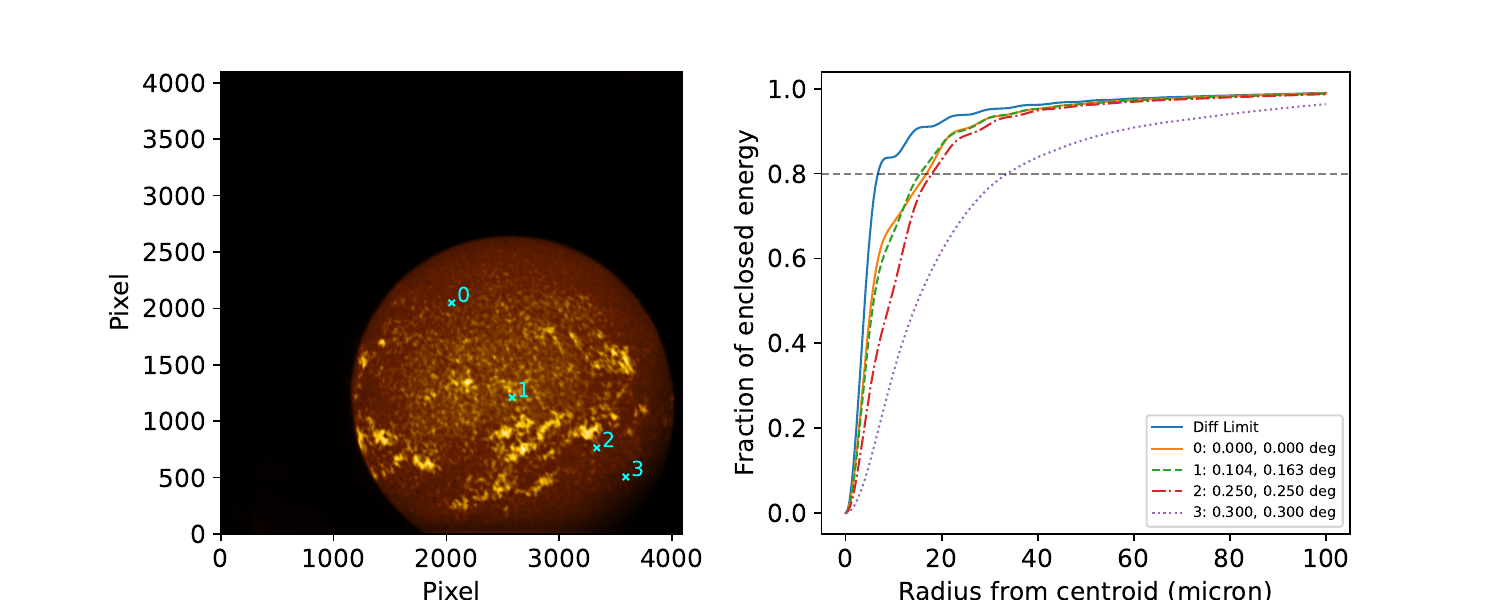}
    \caption{Simulated encircled energy plots for different field points in the \suit\ field of view, and the right panel shows the corresponding encircled energy plots. The left panel shows an image of the sun in the NB03 filter with 4 field points marked. Field point 0: The center of \suit\ CCD (80\% Enclosed energy radius ($ESR_{80}$)= $17 \mu m$), 1: The center of the Sun ($ESR_{80}$= $15.6 \mu m$), 2: field point at $0.250 \times -0.250$ deg from CCD center ($ESR_{80}$= $18 \mu m$), and 3: field point at $0.300 \times -0.300$ deg from CCD center ($ESR_{80}$= $33.8 \mu m$). Notice that the curve of growth for the PSF at field points 0, 1, and 2 are similar, while that for field point 3 is drastically poor.} \label{fig:encircled_energy}
\end{figure}

Due to an unforeseen misalignment between \suit\ and its sister telescope, namely the Visible Emission-Line Coronagraph (VELC), the Sun's image falls on one corner of the \suit\ FOV (see left panel Fig.~\ref{fig:encircled_energy}), while the solar disk is being occulted by VELC. This affects the field of view, as a fraction of the solar disk falls out of the frame, and a significant portion is affected by distortion and comparatively inferior PSF at these high field angles. 

In Fig.~\ref{fig:encircled_energy}, we plot the simulated PSF at four field points marked on the left panel with the corresponding enclosed energy curves for each point on the right panel. The radius enclosing 80\% of the PSF energy is used as the metric to evaluate the sharpness of the PSF. From the figure, it is clear that the radius of the PSF increases with increasing field angle, with that at point three being drastically poorer as expected.

The effect of the PSF on the imaging can be demonstrated with off-pointed observations. Fig.~\ref{fig:corner-dist} shows one such scenario. The same active region was imaged $\sim$ {\bf 16 hours} apart, with the Sun falling on the corner and center of the {\suit} CCD shown in Fig.~\ref{fig:corner-dist}~a and Fig.~\ref{fig:corner-dist}~b, respectively. The full-frame observation clearly shows that the solar disk is distorted and considerably bigger when the Sun is at the CCD's corner. 
For further demonstration, we take AIA 1600~{\AA} observations, co-align, and co-register them to compare with the {\suit} observations. We mark the same active region in panels a \& b with a blue solid line. The SUIT NB03 (solid blue) and AIA 1600~{\AA} (solid black) intensity profiles across the lines are shown in Fig.~\ref{fig:corner-dist}~c and Fig.~\ref{fig:corner-dist}~d, respectively. The SUIT NB4 and AIA 1600~{\AA} intensity profiles match closely when the Sun is pointed at the CCD centre (see Fig.~\ref{fig:corner-dist}~d). 
The limb falls off at a similar angular radius as observed from both {\suit} NB03 and AIA 1600~{\AA}. In contrast, the intensity profile across the same region is stretched out when the Sun falls on the CCD corner and does not agree with the AIA 1600~{\AA} observation. The sharp drop of intensity is not observed in {\suit} NB03, as it slowly decays off (see Fig.~\ref{fig:corner-dist}~c). This stretching of the intensity profiles is a clear indication of diagonal stretching of the image due to distortion at the corner most positions of the CCD.

\subsection{Dark, Bias and Read Noise}
The dark signal refers to the photoelectrons generated by an image sensor per unit time in the absence of any light. For this measurement, we take a single 240~s exposure with a closed entrance aperture and compute the exposure-normalized mean counts in each quadrant. The measured mean count was $1.82 e^-/px/s$, which is significantly better than the specified requirement, i.e., $< 10 e^-/px/s$. 

Similar to the on ground test, we measured the bias and read noise using overscan pixels generated by CCD electronics. 
Our measurements show a read noise of 9 $e^-$ with a bias level of ($466 \pm 0.12\%$) counts, which is in close proximity of the desired read noise of $<$10 photoelectrons per pixel, with a bias value of $<$500 counts. 

\subsection{Flare Localization and auto exposure}
The telescope is equipped with an automatic flare detection module which helps to automatically detect, localize and perform high cadence imaging of an emerging solar flare. While the telescope has its internal trigger module that continuously monitors the increase in intensity in the \suit\ images, the provision is also made to get a flare flag from the Solar Low Energy X-ray Spectrometer (SoLEXS). To generate internal triggers, we have a parallel stream of onboard data generation where the instrument takes 2k$\times$2k binned images using Mg~\rm{II}~k (NB03 filter) every minute. Moreover, we obtain a parallel data stream from the High Energy L1 Orbiting Spectrometer (HEL1OS) onboard Aditya-L1 to generate a flare flag. For further details of the algorithms, see \cite{suit_algo}. 

As soon as a trigger is received, \suit\ switches to flare mode, wherein the telescope takes four images within a minute in the Mg~\rm{II}~k filter to localize the event. If the onboard logic can isolate the flaring region, SUIT stops full disk observations and performs Region of Interest (RoI) imaging. This is combined with automated exposure control, such that the increasing brightness of the flare does not saturate the pixels \citep{suit_algo}.

As the \suit\ CCD has four quadrants which are read independently \citep{suit_main}, the cadence of the observation depends on the position of the RoI on the CCD. 
For observation with a single filter, the cadence of observation ranges from 1.39 s when the RoI is at the CCD center with some portion of the RoI being read by each of the 4 quadrants, to 3.08 sec when the RoI completely lies in one quadrant and is read through one readout.
In the normal flare mode, observations are taken in all eleven filters, and the cadence is limited by the time taken to cycle through the filters. 

\begin{figure}
    \centering
    \includegraphics[width=1\linewidth]{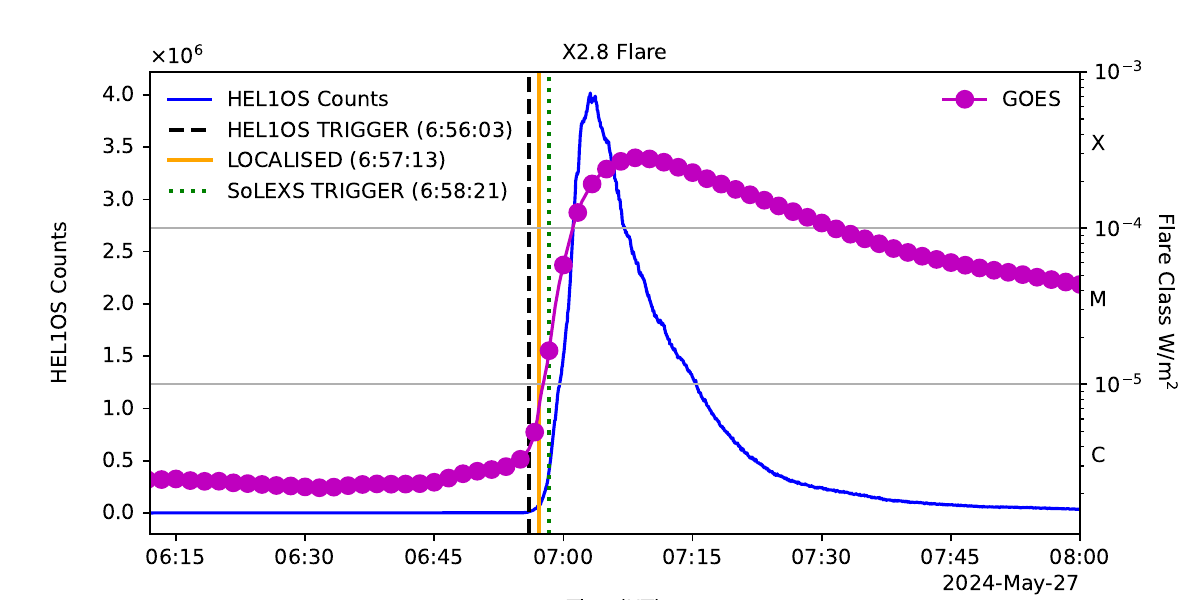}
    \includegraphics[width=1\linewidth]{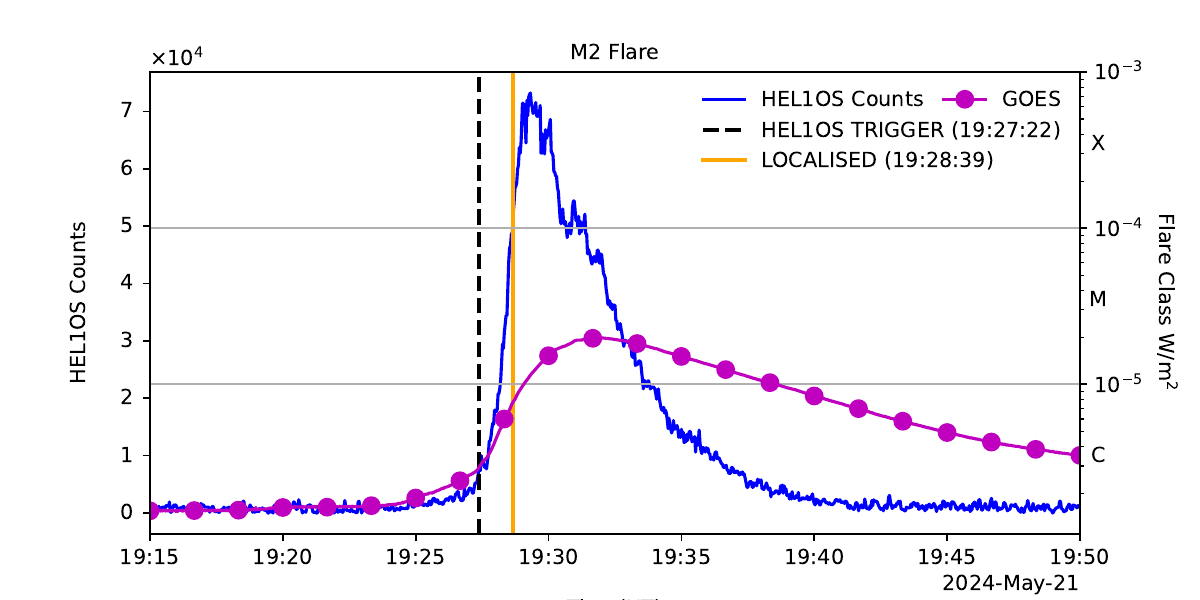}    
    \caption{\js{The figures in the top and bottom panels show} the X-ray light curves from HEL1OS (blue curve) and GOES (pink dot-dash curve). The times for HEL1OS trigger (black dashed vertical line), SoLEXS trigger (green dotted line), and the time of flare localization by \suit\ (yellow solid vertical line) \js{are shown for an X2.8 flare [top] and an M2 flare [bottom]}.}
    \label{fig:flare_detection}
\end{figure}

Since deployment of the flare detection algorithm, there have been 179 M class or stronger flares till July 31, 2024 while {\suit} was observing. 13 of these were off-limb flares, with insufficient NUV signal. Out of the remaining 163 flares, the flare detection algorithm has localized 55 flares. Note that these numbers are from the payload verification phase when several calibration activities were ongoing for \suit, as well as six other payloads on the spacecraft. Moreover, as mentioned earlier, due to the misalignment with VELC, the telescope is almost blind for about 1/3rd of the disk as far as flare localization is concerned. From observations so far, the best identification is achieved by creating a flare flag with HEL1OS data. \js{Fig. \ref{fig:flare_detection} represents an X2.8 flare [top panel] that occurred on May 27, 2024 and an M2 flare [bottom panel] that occurred on May 21, 2024. The blue solid plot is the hard X-ray light curve detected by HEL1OS. The pink dot-dashed line shows the GOES flux for the same flare. 
The vertical lines denote 1. the time stamps for the flare trigger generated on-board {\suit} from HEL1OS data (black-dashed line), 2. the time of flare localization (yellow-solid line), 3. and the flare flag given by SoLEXS (green-dotted line). We notice that {\suit} successfully localized the flare and automatically switched to flare mode, with the high-cadence region of interest observation, within $ \sim 1$ min of flare detection from HEL1OS data for both X2.8 and M2 flares. This demonstrates the sensitivity of the onboard flare detection algorithm over a wide energy range.}

\section{On-board Calibration-Level 1 data}\label{s:calibration}

\subsection{Gain Correction}
The Gain coefficient defines the number of photoelectrons required to register one Analog-to-Digital Unit (ADU) by the Analog-to-Digital Converter (ADC). The \suit\ CCD has four quadrants for which the gain values are 3.04, 3.09, 3.01, and 2.95, respectively. To normalize the image to the same gain settings, each image quadrant is multiplied by the corresponding gain to convert the counts to photoelectrons. This is then divided by the mean gain for all quadrants to get a normalized image representing the signal in counts. 

\subsection{Pixel Response Non Uniformity (PRNU)}
The Pixel Response Non-Uniformity (PRNU) refers to the pixel-to-pixel variation in sensitivity. The gain of every pixel in an image sensor is not the same. This leads to a variation of sensitivity across the image. To mitigate this issue, the image sensor is illuminated with a uniform light source, and the relative variation in sensitivity is measured from pixel to pixel. 
    
SUIT has a collection of 16 LEDs, 8 each for wavelength bands, namely 355~nm and 255~nm. For each wavelength, 4 LEDs of the same wavelength glow simultaneously to illuminate the CCD. This is used to measure the PRNU of the CCD.

The LED image has a large scale illumination pattern, and a pixel-scale fluctuation due to the PRNU. The large-scale illumination pattern (order of 50 pixels) is isolated by boxcar blurring and removed from the LED image. With the large-scale patterns removed, the residual gives the pixel-to-pixel variation in sensitivity, which is the PRNU profile. We divide this profile from Level 0 images to make the necessary corrections. The residual photometric errors after performing PRNU correction are $\sim 0.55\%$ and $\sim 0.36\%$ at 355 nm and 255 nm, respectively.

\subsection{Flat Field}
The Flat Field of a telescope system records the non-uniformity in illumination across the field of view (FOV). While the PRNU records the small-scale pixel-to-pixel response variation of the image sensor, the flat field is mainly used to characterize intensity variations over scales of several tens of pixels and higher.

\begin{figure}
        \centering
        \includegraphics[width=1\linewidth]{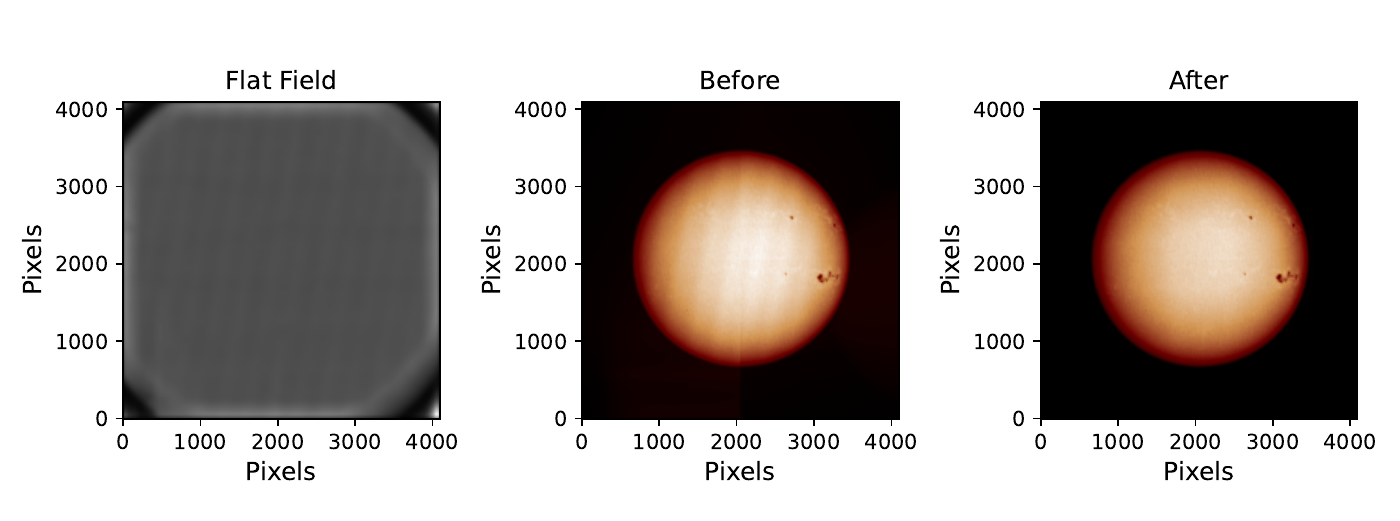}
        \caption{Flat Field correction to remove diagonal stripe-like pattern in \suit\ images. From left to right, the panels show the generated flat field characterizing the stripe-like pattern, the middle panel shows an uncorrected image in the BB03 filter, and the right panel shows the flat field corrected version.}
        \label{fig:flat_field}
    \end{figure}

These flat field images characterize the unevenness of illumination. Generally, the telescope aperture is exposed to a uniform intensity of light, and images are recorded to characterize the non-uniformity. For \suit, a novel method was developed to generate the flat field from several off-pointed solar images combined to illuminate the field of view (FOV). The satellite is off-pointed in steps of 8 arc minutes in vertical and horizontal axes to cover the complete field of view with the Sun as per the scheme in Fig. \ref{fig:off-pointing}. These images are combined using a maximization algorithm, which maps the highest values of all pixels in a stack of images to one image. A detailed account of this method is given in \S\ref{sec:ob_fov}. Boxcar convolution removes the small-scale structures from this image, preserving the large-scale illumination pattern. The generated flat field is divided from the image to apply the correction as shown in Fig. \ref{fig:flat_field}. The achieved intensity variation across the complete field of view is $< 0.11\%$. The residual photometric error after flat field and PRNU correction is 0.56\%, which meets the requirement of $< 1\%$.

\subsection{Scatter Correction}\label{sec:scatter}
The \suit\ telescope is susceptible to scattered sunlight. Due to the high intensity of the Sun, some portion of the sunlight entering the telescope cavity is scattered, eventually reaching the CCD. This scattered light needs to be modeled and removed from the images to make reliable observations.

\begin{figure}[h!]
\centering
\includegraphics[width=12cm]{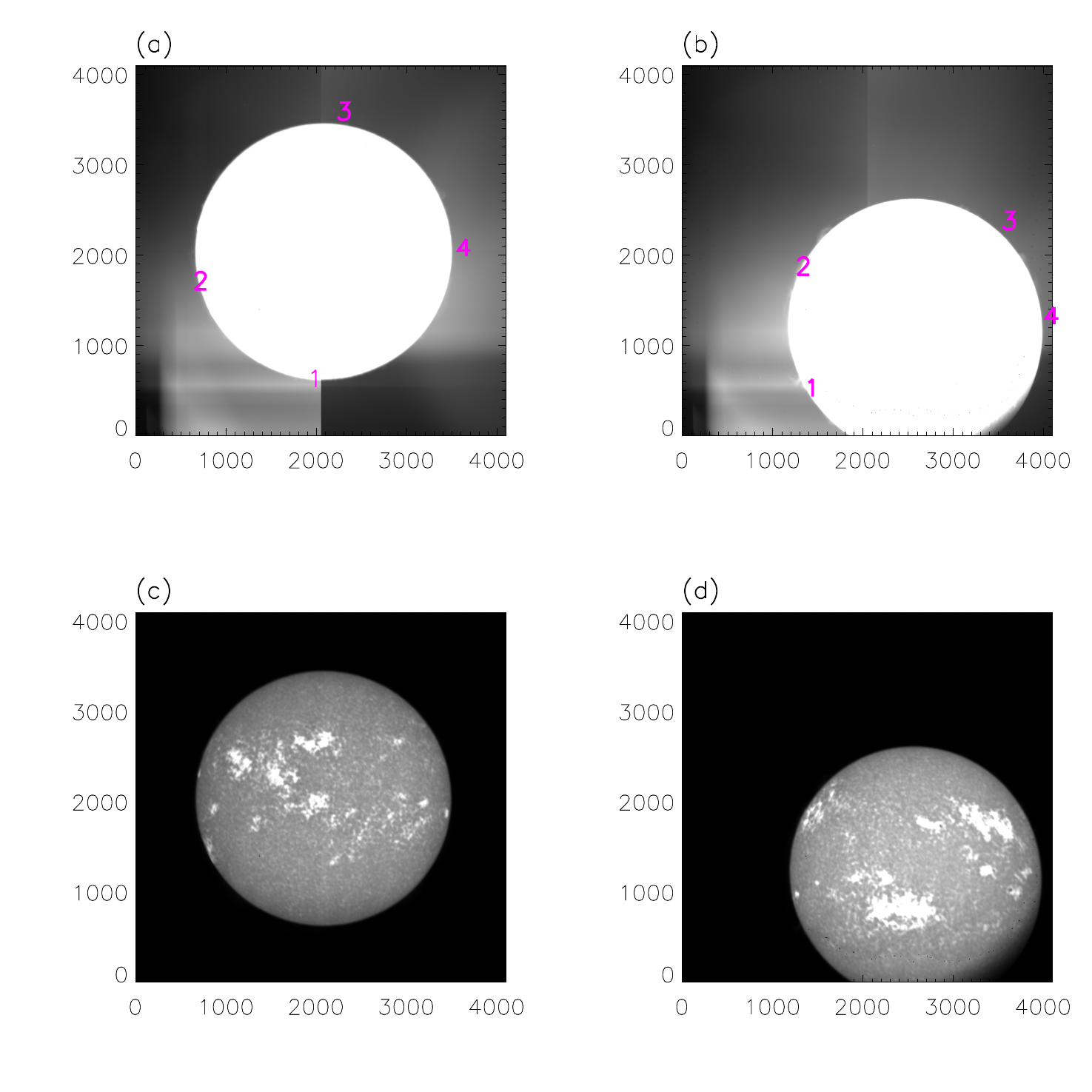}
\caption{Panel (a): NB04 filter image when the Sun is at the center of the CCD. Image intensity is scaled so that the scatter is visible compared to the Sun's intensity. Panel (b): NB04 image at VELC aligned position. Scatter-corrected versions of images in panels (a) and (b) are shown in panels (c) and (d), respectively.}
\label {fig_sct}
\end{figure}

Fig. \ref{fig_sct} shows the bias and gain corrected image of the Sun in the NB04 filter at two different locations of the CCD. Panel (a) is the SUIT-centered observation while panel (b) is the SUIT image at VELC aligned position. Both images are scaled in such a way that the background pattern is visible compared to the solar intensity. Both the images show a background pattern which appears due to the scattered light falling onto the CCD. 
Note that there is a sharp horizontal jump in the scatter pattern at the quadrant boundary, but not in the vertical direction. This can be traced back to how the signal is readout from the \suit\ CCD. In \suit\ full-disk images, the readout happens in the horizontal direction. Therefore, the signal due to scattered light accumulates during the readout for pixels closer to the center, creating a sharp vertical jump at the quadrant boundary. Furthermore, the observed background pattern is filter-dependent, and it varies when the position of the Sun changes significantly on the CCD.
For calibration, we have to find the background pattern that extends into the solar disk region and subtract it from the images. The scatter model for each filter was generated using the off-pointed observations of the Sun. The details of the scatter correction method for the SUIT-centered observations are given below,
\begin{itemize}
\item{We mask the solar disk and the surrounding limb region at a radius below 1600 pixels from the center of the Sun. Everything above this radius is considered as background pattern and not of solar origin.
The typical radius of the Sun is $\sim$1400 pixels.}
\item{Then, \js{identify} two off-pointed images of the Sun where the observed scatter pattern appears similar to the SUIT-centered observation. Furthermore, the background pattern of the off-pointed image should overlap with the masked-out region of the SUIT-centered observation.}
\item{In the next step, we crop the scatter pattern of the off-pointed image and replaced it at the corresponding location in the masked-out region. Then scaled it to match the background of the masked image.}
\item{The scaling is done separately for each horizontal slice of pixels. Each slice has two calculated scaling factors from the left and right limbs of the masked region. We linearly interpolate those values to find the scaling factor for the remaining pixels of the horizontal slice.}
\item{Then, we repeat the above process for each horizontal slice and create the scaled scatter pattern that matches the background of the SUIT center image. Then, the modeled scatter is used to remove the background pattern from the images.}
\end{itemize}

The scatter-corrected images are presented in the bottom panel of \ref{fig_sct}.  A quantitative evaluation of the scatter correction method is presented in Table \ref{tab_sct}, showing the residual due to scatter is typically well below 1\%, except in Ca~\rm{II}, due to the presence of a ghost.

\begin{table}[h!]
\centering
\begin{tabular}{c c cc cc cc c} 
 \hline
 \textbf{Filter} & \multicolumn{2}{c}{\textbf{\% Scatter at 1}} & \multicolumn{2}{c}{\textbf{\% Scatter at 2}} & \multicolumn{2}{c}{\textbf{\% Scatter at 3}} & \multicolumn{2}{c}{\textbf{\% Scatter at 4}} \\
   & \textbf{Before} &  \textbf{After} & \textbf{Before}  & \textbf{After}  & \textbf{Before}  & \textbf{After} & \textbf{Before} & \textbf{After} \\ 
     \hline
     NB01 &   10.96 & 0.55 & 10.42 & 0.46 & 6.22 & 0.20 & 6.27 & 0.39  \\ 
     NB02 &  5.33 & 0.57 & 4.95 & 0.22 & 4.01 & 0.15 & 4.66 & 0.35 \\
     NB03 &  19.15 & 1.36 & 17.83 & 1.17 & 12.68 & 0.59 & 14.00 & 0.99 \\
     NB04 &  14.31 & 0.59 & 14.12 & 0.92 & 10.13 & 0.37 & 11.34 & 0.74 \\
     NB05 &  5.55 & 0.29 & 5.54 & 0.27 & 4.25 & 0.24 & 5.30 & 0.65 \\ 
     NB06 &  5.98 & 0.26 & 5.39 & 0.24 & 3.96 & 0.029 & 3.47 & 0.18  \\ 
     NB07 &   5.07 & 0.26 & 4.50 & 0.16 & 2.71 & 0.27 & 3.28  & 1.33 \\ 
     NB08 &  27.87  & 3.82 & 32.38 & 7.09 & 34.46 & 14.66 & 26.40 & 6.23  \\ 
     BB01 &  6.30 & 0.49 & 6.10 & 0.46 & 4.13 & 0.28 & 6.57 & 1.12  \\ 
     BB02 & 8.39 & 0.12 & 8.12 & 0.18 & 11.59 & 0.31 & 9.45 & 0.26 \\ 
     BB03 &  6.60 & 0.21 & 5.85 & 0.21 & 3.53 & 0.12 & 6.03 & 3.07  \\ 
     \hline
\end{tabular}
\caption{Comparison of scatter level (in \%) with respect to the mean sun center intensity before and after the scatter subtraction for VELC aligned position. The location chosen for scatter measurement is shown in panel (b) of Fig. \ref{fig_sct}.}\label{tab_sct}
\end{table}

\subsection{Ghost Correction}

\begin{figure}
    \centering
    \includegraphics[width=1.0\linewidth]{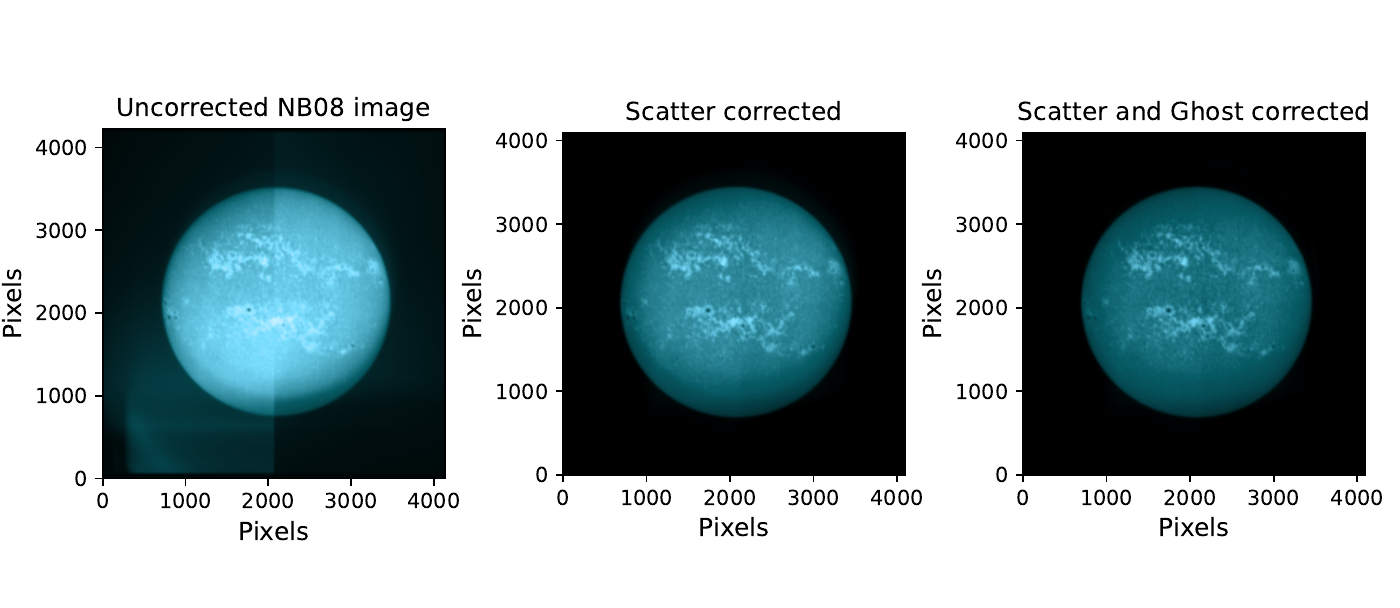}
    \caption{Scatter and Ghost correction applied on \suit\ NB08 image. The left panel shows an uncorrected NB08 image from May 17, 2024. The middle panel shows the same image after removing the contribution from scattered light. The right panel shows the same image after removing scattered light and ghost reflections.}
    \label{fig:scatter_ghost}
\end{figure}

{\suit} operates with a stack of two filters required to pass the relevant band of light. The filters, being dichroic in nature, have highly reflective surfaces. This leads to several reflections between the filters. To minimize this issue, all {\suit} science filters are tilted at specific angles, except NB08 (Ca~\rm{II}). Hence, the Ca~\rm{II} images are affected by ghost reflections, as shown in Fig.~\ref{fig:scatter_ghost}. Panel A is the uncorrected image, which contains scattered light and ghost reflection. The ghost reflection can be seen clearly after scatter correction in the image shown in panel B. Due to the slight shift in ghost reflection, a portion of the solar disk is ghost-free and it appears darker compared to the rest of the solar disk (See left solar limb in panel B). 
Furthermore, the shift also creates a region that has only ghost reflection without any solar features. Based on these two regions, we find that the ghost is 8.2 times dimmer than the corresponding signal. Then, we remove the ghost next to the ghost-free region of the solar disk using the assumption that the overlying part of the ghost is also 8.2 times dimmer than the ghost-free signal next to it. This iteration is repeated until the ghost is completely eliminated from the solar disk. The cumulative intensity of scattered light and ghost reflection for NB08 filter images is $\sim 28\% $. This is reduced to $1.67\%$ after applying scatter correction (\ref{sec:scatter}) and ghost correction as illustrated in Fig.~\ref{fig:scatter_ghost}.

\begin{table}
    \centering
    \begin{tabular}{cccc}
    \hline
        \textbf{Parameter} & \textbf{Specification} & \textbf{Ground} & \textbf{Onboard} \\
        & & \textbf{Test Result} & \textbf{Test Result} \\
        \hline
        Field of View & $0.39^\circ$ radius & $0.39^\circ$ radius & $0.39^\circ$ radius\\
        PSF 80\% EE & $1.4"$ & $2.2"$ at NB04 & TBT \footnotemark[1]\\
        Plate Scale & $0.7"/px$ & $0.69"/px$ & $0.69"/px$\\
        MTF & 10\% at 42lpmm & 10\% at 23 lpmm & TBT \footnotemark[1]\\
        Read Noise & $< 10 e^-$ & $8 e^-$ & $9e^-$ \\
        Bias & $<500~ADU$ & $\sim 2500~ADU$ & $466~ADU \pm 0.12 \%$\\
        Mean Dark Signal & $<10~e^-/px/s$ &  & $1.82~e^-/px/s$\\
        PRNU (Post Correction) & < 1\% & & $\sim 0.43\%$ at 355 nm\\
        &  &  & $\sim 0.33\%$ at 255 nm  \\
        Flat Field & < 1\% &  & < 0.11\%\\
        (Post Correction) &  &  & \\
        Total Photometric Error & < 1\% &  & 0.56\%\\
        (Post Correction) &  &  & \\
        \hline
        \footnotetext[1]{To be tested}
    \end{tabular}
    \caption{Calibration tests performed on \suit. The columns indicate the test performed, the requirements as per design, the ground test results and the on board test results.}
    \label{tab:specs}
\end{table}

\section{Summary and conclusions} \label{s:conclusion}
In this paper, we have discussed various on-ground and on-board tests performed on SUIT. The results obtained are shown in Table ~\ref{tab:specs}. In the table, we compare the results obtained from the ground and onboard tests with those specified in the design. We find that all the test results are compliant, except for the spatial resolution and the unfortunate misalignment between \suit\ and VELC. Due to the difference in the specified and measured PSF, the photometry at a scale of 1.4" will not be possible. Instead, it will only be possible at a scale of 2.2". The misalignment between the VELC and \suit\ leads to an unfortunate scenario that will impact the science related to solar spectral irradiance. 
Given the test results, it is demonstrated that SUIT opens up a new window for solar observations.

\begin{acks}
\js{We thank the reviewer for the constructive comments and suggestions}. {\suit} is built by a consortium led by the Inter-University Centre for Astronomy and Astrophysics (IUCAA), Pune, and supported by ISRO as part of the Aditya-L1 mission. The consortium consists of SAG/URSC, MAHE, CESSI-IISER Kolkata (MoE), IIA, MPS, USO/PRL, and Tezpur University. Aditya-L1 is an observatory class mission which is funded and operated by the Indian Space Research Organization. The mission was conceived and realised with the help from all ISRO Centres and payloads were realised by the payload PI Institutes in close collaboration with ISRO and many other national institutes - Indian Institute of Astrophysics (IIA); Inter-University Centre of Astronomy and Astrophysics (IUCAA); Laboratory for Electro-optics System (LEOS) of ISRO; Physical Research Laboratory (PRL); U R Rao Satellite Centre of ISRO; Vikram Sarabhai Space Centre (VSSC) of ISRO. The AIA and HMI data used here are courtesy of SDO (NASA) and the AIA and HMI consortium. SKS acknowledges funding from the European Research Council (ERC) under the European Union's Horizon 2020 research and innovation programme (grant agreement No. 101097844 — project WINSUN). This research used version 4.1.5 \citep{sunpy_ver} of the SunPy open-source software package \citep{sunpy20} and PYTHON packages NumPy \citep{numpy}, Matplotlib \citep{matpltolib}, SciPy \citep{scipy}, Photutils, an Astropy package for
detection and photometry of astronomical sources \citep{photutils} and SciencePlots \citep{SciencePlots}.
\end{acks}

%
%
%
%
%
%
%


\bibliographystyle{spr-mp-sola}
\bibliography{first_review_main.bib}

\begin{thebibliography}{19}
\ifx\bisbn     \undefined \def\bisbn  #1{ISBN #1}\fi
\ifx\binits    \undefined \def\binits#1{#1}\fi
\ifx\bauthor   \undefined \def\bauthor#1{#1}\fi
\ifx\batitle   \undefined \def\batitle#1{#1}\fi
\ifx\bjtitle   \undefined \def\bjtitle#1{\textit{#1}}\fi
\ifx\bvolume   \undefined \def\bvolume#1{\textbf{#1}}\fi
\ifx\byear     \undefined \def\byear#1{#1}\fi
\ifx\bissue    \undefined \def\bissue#1{#1}\fi
\ifx\bfpage    \undefined \def\bfpage#1{#1}\fi
\ifx\blpage    \undefined \def\blpage #1{#1}\fi
\ifx\burl      \undefined \def\burl#1{\href{#1}{\textsf{URL}}}\fi
\ifx\href      \undefined \def\href#1#2{#2}\fi
\ifx\betal     \undefined \def\betal{et al.}\fi
\ifx\bctitle   \undefined \def\bctitle#1{#1}\fi
\ifx\beditor   \undefined \def\beditor#1{#1}\fi
\ifx\bbtitle   \undefined \def\bbtitle#1{\textit{#1}}\fi
\ifx\bedition  \undefined \def\bedition#1{#1}\fi
\ifx\bseriesno \undefined \def\bseriesno#1{\textbf{#1}}\fi
\ifx\blocation \undefined \def\blocation#1{#1}\fi
\ifx\bsertitle \undefined \def\bsertitle#1{\textit{#1}}\fi
\ifx\bsnm      \undefined \def\bsnm#1{#1}\fi
\ifx\bsuffix   \undefined \def\bsuffix#1{#1}\fi
\ifx\bparticle \undefined \def\bparticle#1{#1}\fi
\ifx\barticle  \undefined \def\barticle#1{}\fi
\ifx\binstitute  \undefined \def\binstitute#1{#1}\fi
\ifx\bpublisher  \undefined \def\bpublisher#1{#1}\fi
\ifx\doiurl    \undefined \def\doiurl#1{\href{#1}{DOI}}\fi
\makeatletter
\def\safeHref#1#2#3{\in@{http}{#2}\ifin@\href{#2}{#3}\else\href{#1#2}{#3}\fi}
\makeatother
\ifx\adsurl    \undefined \def\adsurl#1{\safeHref{https://ui.adsabs.harvard.edu/abs/}{#1}{ADS}}\fi
\ifx\arxivurl  \undefined \def\arxivurl#1{\safeHref{http://arxiv.org/abs/}{#1}{arXiv}}\fi
\ifx\botherref \undefined \def\botherref#1{}\fi
\ifx\url       \undefined \def\url#1{#1}\fi
\ifx\bchapter  \undefined \def\bchapter#1{}\fi
\ifx\bbook     \undefined \def\bbook#1{}\fi
\ifx\bcomment  \undefined \def\bcomment#1{#1}\fi
\ifx\oauthor   \undefined \def\oauthor#1{#1}\fi
\ifx\citeauthoryear \undefined\def \citeauthoryear#1{#1}\fi
\def\endbibitem {}
\ifx\bconflocation  \undefined \def\bconflocation#1{#1} \fi

\bibitem[\protect\citeauthoryear{Bradley et~al.}{2024}]{photutils}
\begin{botherref}
\oauthor{\bsnm{Bradley}, \binits{L.}},
\oauthor{\bsnm{Sipőcz}, \binits{B.}},
\oauthor{\bsnm{Robitaille}, \binits{T.}},
\oauthor{\bsnm{Tollerud}, \binits{E.}},
\oauthor{\bsnm{Vinícius}, \binits{Z.}},
\oauthor{\bsnm{Deil}, \binits{C.}},
\oauthor{\bsnm{Barbary}, \binits{K.}},
\oauthor{\bsnm{Wilson}, \binits{T.J.}},
\oauthor{\bsnm{Busko}, \binits{I.}},
\oauthor{\bsnm{Donath}, \binits{A.}},
\oauthor{\bsnm{Günther}, \binits{H.M.}},
\oauthor{\bsnm{Cara}, \binits{M.}},
\oauthor{\bsnm{Lim}, \binits{P.L.}},
\oauthor{\bsnm{Meßlinger}, \binits{S.}},
\oauthor{\bsnm{Conseil}, \binits{S.}},
\oauthor{\bsnm{Burnett}, \binits{Z.}},
\oauthor{\bsnm{Bostroem}, \binits{A.}},
\oauthor{\bsnm{Droettboom}, \binits{M.}},
\oauthor{\bsnm{Bray}, \binits{E.M.}},
\oauthor{\bsnm{Bratholm}, \binits{L.A.}},
\oauthor{\bsnm{Ginsburg}, \binits{A.}},
\oauthor{\bsnm{Jamieson}, \binits{W.}},
\oauthor{\bsnm{Barentsen}, \binits{G.}},
\oauthor{\bsnm{Craig}, \binits{M.}},
\oauthor{\bsnm{Morris}, \binits{B.M.}},
\oauthor{\bsnm{Perrin}, \binits{M.}},
\oauthor{\bsnm{Rathi}, \binits{S.}},
\oauthor{\bsnm{Pascual}, \binits{S.}},
\oauthor{\bsnm{Georgiev}, \binits{I.Y.}}:
2024,
astropy/photutils: 2.0.2,
Zenodo.
\doiurl{https://doi.org/10.5281/zenodo.13989456}.
\burl{https://doi.org/10.5281/zenodo.13989456}.
\end{botherref}
\endbibitem

\bibitem[\protect\citeauthoryear{Garrett}{2021}]{SciencePlots}
\begin{botherref}
\oauthor{\bsnm{Garrett}, \binits{J.D.}}:
2021,
{garrettj403/SciencePlots}.
\doiurl{https://doi.org/10.5281/zenodo.4106649}.
\burl{http://doi.org/10.5281/zenodo.4106649}.
\end{botherref}
\endbibitem

\bibitem[\protect\citeauthoryear{{Ghosh} et~al.}{2022}]{thermal}
\begin{bchapter}
\bauthor{\bsnm{{Ghosh}}, \binits{A.}},
\bauthor{\bsnm{{Kesharwani}}, \binits{R.}},
\bauthor{\bsnm{{Khan}}, \binits{A.R.}},
\bauthor{\bsnm{{Padinhatteeri}}, \binits{S.}},
\bauthor{\bsnm{{Tripathi}}, \binits{D.}},
\bauthor{\bsnm{{Ramaprakash}}, \binits{A.N.}},
\bauthor{\bsnm{{Patel}}, \binits{K.}},
\bauthor{\bsnm{{Jalluri}}, \binits{T.D.P.V.}},
\bauthor{\bsnm{{Madhumalathi}}, \binits{G.R.}},
\bauthor{\bsnm{{Venkateshwaran}}, \binits{R.}},
\bauthor{\bsnm{{Elumalai}}, \binits{S.}},
\bauthor{\bsnm{{Gupta}}, \binits{K.}},
\bauthor{\bsnm{{Nair}}, \binits{J.P.}},
\bauthor{\bsnm{{Sparrow}}, \binits{H.}},
\bauthor{\bsnm{{Worlikar}}, \binits{R.S.}},
\bauthor{\bsnm{{Gupta}}, \binits{A.}},
\bauthor{\bsnm{{Sen}}, \binits{S.}}:
\byear{2022},
\bctitle{{The thermal filter for the Solar Ultraviolet Imaging Telescope (SUIT) on-board Aditya-L1}}.
In: \beditor{\bsnm{{den Herder}}, \binits{J.-W.A.}},
\beditor{\bsnm{{Nikzad}}, \binits{S.}},
\beditor{\bsnm{{Nakazawa}}, \binits{K.}} (eds.)
\bbtitle{Space Telescopes and Instrumentation 2022: Ultraviolet to Gamma Ray},
\bsertitle{Society of Photo-Optical Instrumentation Engineers (SPIE) Conference Series}
\bseriesno{12181},
\bfpage{121813O}.
\doiurl{https://doi.org/10.1117/12.2629760}.
\adsurl{2022SPIE12181E..3OG}.
\end{bchapter}
\endbibitem

\bibitem[\protect\citeauthoryear{Harris et~al.}{2020}]{numpy}
\begin{barticle}
\bauthor{\bsnm{Harris}, \binits{C.R.}},
\bauthor{\bsnm{Millman}, \binits{K.J.}},
\bauthor{\bparticle{van~der} \bsnm{Walt}, \binits{S.J.}},
\bauthor{\bsnm{Gommers}, \binits{R.}},
\bauthor{\bsnm{Virtanen}, \binits{P.}},
\bauthor{\bsnm{Cournapeau}, \binits{D.}},
\bauthor{\bsnm{Wieser}, \binits{E.}},
\bauthor{\bsnm{Taylor}, \binits{J.}},
\bauthor{\bsnm{Berg}, \binits{S.}},
\bauthor{\bsnm{Smith}, \binits{N.J.}},
\bauthor{\bsnm{Kern}, \binits{R.}},
\bauthor{\bsnm{Picus}, \binits{M.}},
\bauthor{\bsnm{Hoyer}, \binits{S.}},
\bauthor{\bparticle{van} \bsnm{Kerkwijk}, \binits{M.H.}},
\bauthor{\bsnm{Brett}, \binits{M.}},
\bauthor{\bsnm{Haldane}, \binits{A.}},
\bauthor{\bparticle{del} \bsnm{R{\'{i}}o}, \binits{J.F.}},
\bauthor{\bsnm{Wiebe}, \binits{M.}},
\bauthor{\bsnm{Peterson}, \binits{P.}},
\bauthor{\bsnm{G{\'{e}}rard-Marchant}, \binits{P.}},
\bauthor{\bsnm{Sheppard}, \binits{K.}},
\bauthor{\bsnm{Reddy}, \binits{T.}},
\bauthor{\bsnm{Weckesser}, \binits{W.}},
\bauthor{\bsnm{Abbasi}, \binits{H.}},
\bauthor{\bsnm{Gohlke}, \binits{C.}},
\bauthor{\bsnm{Oliphant}, \binits{T.E.}}:
\byear{2020},
\batitle{Array programming with {NumPy}}.
\bjtitle{Nature}
\bvolume{585},
\bfpage{357}.
\doiurl{https://doi.org/10.1038/s41586-020-2649-2}.
\burl{https://doi.org/10.1038/s41586-020-2649-2}.
\end{barticle}
\endbibitem

\bibitem[\protect\citeauthoryear{Hunter}{2007}]{matpltolib}
\begin{barticle}
\bauthor{\bsnm{Hunter}, \binits{J.D.}}:
\byear{2007},
\batitle{Matplotlib: A 2D graphics environment}.
\bjtitle{Computing in Science \& Engineering}
\bvolume{9},
\bfpage{90}.
\doiurl{https://doi.org/10.1109/MCSE.2007.55}.
\end{barticle}
\endbibitem

\bibitem[\protect\citeauthoryear{{Mcclintock}, {Rottman}, and {Woods}}{2005}]{mcclintock05}
\begin{barticle}
\bauthor{\bsnm{{Mcclintock}}, \binits{W.E.}},
\bauthor{\bsnm{{Rottman}}, \binits{G.J.}},
\bauthor{\bsnm{{Woods}}, \binits{T.N.}}:
\byear{2005},
\batitle{{Solar-Stellar Irradiance Comparison Experiment II (Solstice II): Instrument Concept and Design}}.
\bjtitle{\solphys}
\bvolume{230},
\bfpage{225}.
\doiurl{https://doi.org/10.1007/s11207-005-7432-x}.
\adsurl{2005SoPh..230..225M}.
\end{barticle}
\endbibitem

\bibitem[\protect\citeauthoryear{{Meftah} et~al.}{2018}]{2018A&A...616A..64M}
\begin{barticle}
\bauthor{\bsnm{{Meftah}}, \binits{M.}},
\bauthor{\bsnm{{Corbard}}, \binits{T.}},
\bauthor{\bsnm{{Hauchecorne}}, \binits{A.}},
\bauthor{\bsnm{{Morand}}, \binits{F.}},
\bauthor{\bsnm{{Ikhlef}}, \binits{R.}},
\bauthor{\bsnm{{Chauvineau}}, \binits{B.}},
\bauthor{\bsnm{{Renaud}}, \binits{C.}},
\bauthor{\bsnm{{Sarkissian}}, \binits{A.}},
\bauthor{\bsnm{{Dam{\'e}}}, \binits{L.}}:
\byear{2018},
\batitle{{Solar radius determined from PICARD/SODISM observations and extremely weak wavelength dependence in the visible and the near-infrared}}.
\bjtitle{\aap}
\bvolume{616},
\bfpage{A64}.
\doiurl{https://doi.org/10.1051/0004-6361/201732159}.
\adsurl{2018A&A...616A..64M}.
\end{barticle}
\endbibitem

\bibitem[\protect\citeauthoryear{Mumford et~al.}{2023}]{sunpy_ver}
\begin{botherref}
\oauthor{\bsnm{Mumford}, \binits{S.J.}},
\oauthor{\bsnm{Freij}, \binits{N.}},
\oauthor{\bsnm{Stansby}, \binits{D.}},
\oauthor{\bsnm{Christe}, \binits{S.}},
\oauthor{\bsnm{Ireland}, \binits{J.}},
\oauthor{\bsnm{Mayer}, \binits{F.}},
\oauthor{\bsnm{Shih}, \binits{A.Y.}},
\oauthor{\bsnm{Hughitt}, \binits{V.K.}},
\oauthor{\bsnm{Ryan}, \binits{D.F.}},
\oauthor{\bsnm{Liedtke}, \binits{S.}},
\oauthor{\bsnm{Hayes}, \binits{L.}},
\oauthor{\bsnm{Pérez-Suárez}, \binits{D.}},
\oauthor{\bsnm{I.}, \binits{V.K.}},
\oauthor{\bsnm{Barnes}, \binits{W.}},
\oauthor{\bsnm{Chakraborty}, \binits{P.}},
\oauthor{\bsnm{Inglis}, \binits{A.}},
\oauthor{\bsnm{Pattnaik}, \binits{P.}},
\oauthor{\bsnm{Sipőcz}, \binits{B.}},
\oauthor{\bsnm{MacBride}, \binits{C.}},
\oauthor{\bsnm{Sharma}, \binits{R.}},
\oauthor{\bsnm{Leonard}, \binits{A.}},
\oauthor{\bsnm{Hewett}, \binits{R.}},
\oauthor{\bsnm{Hamilton}, \binits{A.}},
\oauthor{\bsnm{Manhas}, \binits{A.}},
\oauthor{\bsnm{Panda}, \binits{A.}},
\oauthor{\bsnm{Earnshaw}, \binits{M.}},
\oauthor{\bsnm{Choudhary}, \binits{N.}},
\oauthor{\bsnm{Kumar}, \binits{A.}},
\oauthor{\bsnm{Singh}, \binits{R.}},
\oauthor{\bsnm{Chanda}, \binits{P.}},
\oauthor{\bsnm{Haque}, \binits{M.A.}},
\oauthor{\bsnm{Kirk}, \binits{M.S.}},
\oauthor{\bsnm{Mueller}, \binits{M.}},
\oauthor{\bsnm{Konge}, \binits{S.}},
\oauthor{\bsnm{Srivastava}, \binits{R.}},
\oauthor{\bsnm{Wentzel-Long}, \binits{M.}},
\oauthor{\bsnm{Jain}, \binits{Y.}},
\oauthor{\bsnm{Bennett}, \binits{S.}},
\oauthor{\bsnm{Baruah}, \binits{A.}},
\oauthor{\bsnm{Arbolante}, \binits{Q.}},
\oauthor{\bsnm{Charlton}, \binits{M.}},
\oauthor{\bsnm{Maloney}, \binits{S.}},
\oauthor{\bsnm{Mishra}, \binits{S.}},
\oauthor{\bsnm{Paul}, \binits{J.A.}},
\oauthor{\bsnm{Verma}, \binits{A.}},
\oauthor{\bsnm{Chorley}, \binits{N.}},
\oauthor{\bsnm{Chouhan}, \binits{A.}},
\oauthor{\bsnm{Himanshu}},
\oauthor{\bsnm{Mason}, \binits{J.P.}},
\oauthor{\bsnm{Zivadinovic}, \binits{L.}},
\oauthor{\bsnm{Modi}, \binits{S.}},
\oauthor{\bsnm{Sharma}, \binits{Y.}},
\oauthor{\bsnm{Naman9639}},
\oauthor{\bsnm{Bobra}, \binits{M.G.}},
\oauthor{\bsnm{Rozo}, \binits{J.I.C.}},
\oauthor{\bsnm{Manley}, \binits{L.}},
\oauthor{\bsnm{Ivashkiv}, \binits{K.}},
\oauthor{\bsnm{Laitinen}, \binits{T.}},
\oauthor{\bsnm{Chatterjee}, \binits{A.}},
\oauthor{\bparticle{von} \bsnm{Forstner}, \binits{J.F.}},
\oauthor{\bsnm{Bazán}, \binits{J.}},
\oauthor{\bsnm{Stern}, \binits{K.A.}},
\oauthor{\bsnm{Gieseler}, \binits{J.}},
\oauthor{\bsnm{Evans}, \binits{J.}},
\oauthor{\bsnm{Jain}, \binits{S.}},
\oauthor{\bsnm{Malocha}, \binits{M.}},
\oauthor{\bsnm{Ghosh}, \binits{S.}},
\oauthor{\bsnm{Airmansmith97}},
\oauthor{\bsnm{Stańczak}, \binits{D.}},
\oauthor{\bsnm{Singh}, \binits{R.R.}},
\oauthor{\bsnm{Visscher}, \binits{R.D.}},
\oauthor{\bsnm{Verma}, \binits{S.}},
\oauthor{\bsnm{SophieLemos}},
\oauthor{\bsnm{Agrawal}, \binits{A.}},
\oauthor{\bsnm{Alam}, \binits{A.}},
\oauthor{\bsnm{Buddhika}, \binits{D.}},
\oauthor{\bsnm{Pathak}, \binits{H.}},
\oauthor{\bsnm{Rideout}, \binits{J.R.}},
\oauthor{\bsnm{Sharma}, \binits{S.}},
\oauthor{\bsnm{Park}, \binits{J.}},
\oauthor{\bsnm{Bates}, \binits{M.}},
\oauthor{\bsnm{Wilson}, \binits{A.}},
\oauthor{\bsnm{Shukla}, \binits{D.}},
\oauthor{\bsnm{Giger}, \binits{M.}},
\oauthor{\bsnm{Mishra}, \binits{P.}},
\oauthor{\bsnm{Sharma}, \binits{D.}},
\oauthor{\bsnm{Goel}, \binits{D.}},
\oauthor{\bsnm{Taylor}, \binits{G.}},
\oauthor{\bsnm{Cetusic}, \binits{G.}},
\oauthor{\bsnm{Reiter}, \binits{G.}},
\oauthor{\bsnm{Jacob}},
\oauthor{\bsnm{Inchaurrandieta}, \binits{M.}},
\oauthor{\bsnm{Dacie}, \binits{S.}},
\oauthor{\bsnm{Dubey}, \binits{S.}},
\oauthor{\bsnm{Eigenbrot}, \binits{A.}},
\oauthor{\bsnm{Bray}, \binits{E.M.}},
\oauthor{\bsnm{Surve}, \binits{R.}},
\oauthor{\bsnm{Zahniy}, \binits{S.}},
\oauthor{\bsnm{Sidhu}, \binits{S.}},
\oauthor{\bsnm{Meszaros}, \binits{T.}},
\oauthor{\bsnm{Parkhi}, \binits{U.}},
\oauthor{\bsnm{Russell}, \binits{W.}},
\oauthor{\bsnm{Bose}, \binits{A.}},
\oauthor{\bsnm{Pandey}, \binits{A.}},
\oauthor{\bsnm{Price-Whelan}, \binits{A.}},
\oauthor{\bsnm{J}, \binits{A.}},
\oauthor{\bsnm{Chicrala}, \binits{A.}},
\oauthor{\bsnm{Ankit}},
\oauthor{\bsnm{Guennou}, \binits{C.}},
\oauthor{\bsnm{D'Avella}, \binits{D.}},
\oauthor{\bsnm{Williams}, \binits{D.}},
\oauthor{\bsnm{Verma}, \binits{D.}},
\oauthor{\bsnm{Ballew}, \binits{J.}},
\oauthor{\bsnm{Agrawal}, \binits{K.}},
\oauthor{\bsnm{Murphy}, \binits{N.}},
\oauthor{\bsnm{Lodha}, \binits{P.}},
\oauthor{\bsnm{Robitaille}, \binits{T.}},
\oauthor{\bsnm{Augspurger}, \binits{T.}},
\oauthor{\bsnm{Krishan}, \binits{Y.}},
\oauthor{\bsnm{honey}},
\oauthor{\bsnm{neerajkulk}},
\oauthor{\bsnm{Bhope}, \binits{A.}},
\oauthor{\bsnm{Gaba}, \binits{A.S.}},
\oauthor{\bsnm{Hill}, \binits{A.}},
\oauthor{\bsnm{Mampaey}, \binits{B.}},
\oauthor{\bsnm{Wiedemann}, \binits{B.M.}},
\oauthor{\bsnm{Molina}, \binits{C.}},
\oauthor{\bsnm{Briseno}, \binits{D.G.}},
\oauthor{\bsnm{Keşkek}, \binits{D.}},
\oauthor{\bsnm{Habib}, \binits{I.}},
\oauthor{\bsnm{Letts}, \binits{J.}},
\oauthor{\bsnm{Singaravelan}, \binits{K.}},
\oauthor{\bsnm{Ranjan}, \binits{K.}},
\oauthor{\bsnm{Altunian}, \binits{N.}},
\oauthor{\bsnm{Streicher}, \binits{O.}},
\oauthor{\bsnm{Gomillion}, \binits{R.}},
\oauthor{\bsnm{Agarwal}, \binits{S.}},
\oauthor{\bsnm{Kothari}, \binits{Y.}},
\oauthor{\bsnm{Nomiya}, \binits{Y.}},
\oauthor{\bsnm{mridulpandey}},
\oauthor{\bsnm{Stevens}, \binits{A.L.}},
\oauthor{\bsnm{B}, \binits{A.}},
\oauthor{\bsnm{Bahuleyan}, \binits{A.}},
\oauthor{\bsnm{Kaszynski}, \binits{A.}},
\oauthor{\bsnm{W}, \binits{A.}},
\oauthor{\bsnm{Mehrotra}, \binits{A.}},
\oauthor{\bsnm{Tang}, \binits{A.}},
\oauthor{\bsnm{Sinha}, \binits{A.}},
\oauthor{\bsnm{Smith}, \binits{A.}},
\oauthor{\bsnm{Kustov}, \binits{A.}},
\oauthor{\bsnm{Stone}, \binits{B.}},
\oauthor{\bsnm{Bard}, \binits{C.}},
\oauthor{\bsnm{Arias}, \binits{E.}},
\oauthor{\bsnm{Tollerud}, \binits{E.}},
\oauthor{\bsnm{Dover}, \binits{F.M.}},
\oauthor{\bsnm{Verstringe}, \binits{F.}},
\oauthor{\bsnm{Kumar}, \binits{G.}},
\oauthor{\bsnm{Mathur}, \binits{H.}},
\oauthor{\bsnm{Babuschkin}, \binits{I.}},
\oauthor{\bsnm{Calixto}, \binits{J.}},
\oauthor{\bsnm{Wimbish}, \binits{J.}},
\oauthor{\bsnm{Qing}, \binits{J.}},
\oauthor{\bsnm{Buitrago-Casas}, \binits{J.C.}},
\oauthor{\bsnm{Krishna}, \binits{K.}},
\oauthor{\bsnm{Chaudhari}, \binits{K.}},
\oauthor{\bsnm{Hiware}, \binits{K.}},
\oauthor{\bsnm{Ghosh}, \binits{K.}},
\oauthor{\bsnm{Lyes}, \binits{M.M.}},
\oauthor{\bsnm{Mangaonkar}, \binits{M.}},
\oauthor{\bsnm{Cheung}, \binits{M.}},
\oauthor{\bsnm{Mendero}, \binits{M.}},
\oauthor{\bsnm{Dedhia}, \binits{M.}},
\oauthor{\bsnm{Schoentgen}, \binits{M.}},
\oauthor{\bsnm{Shahdadpuri}, \binits{N.}},
\oauthor{\bsnm{Srinivasan}, \binits{N.}},
\oauthor{\bsnm{Gyenge}, \binits{N.G.}},
\oauthor{\bsnm{Mekala}, \binits{R.R.}},
\oauthor{\bsnm{Das}, \binits{R.}},
\oauthor{\bsnm{Mishra}, \binits{R.}},
\oauthor{\bsnm{Sharma}, \binits{R.}},
\oauthor{\bsnm{Srikanth}, \binits{S.}},
\oauthor{\bsnm{Jain}, \binits{S.}},
\oauthor{\bsnm{Kannojia}, \binits{S.}},
\oauthor{\bsnm{Yadav}, \binits{T.}},
\oauthor{\bsnm{Paul}, \binits{T.}},
\oauthor{\bsnm{Wilkinson}, \binits{T.D.}},
\oauthor{\bsnm{Caswell}, \binits{T.A.}},
\oauthor{\bsnm{Braccia}, \binits{T.}},
\oauthor{\bsnm{Pereira}, \binits{T.M.D.}},
\oauthor{\bsnm{Gates}, \binits{T.}},
\oauthor{\bsnm{Dang}, \binits{T.K.}},
\oauthor{\bsnm{Bankar}, \binits{V.}},
\oauthor{\bsnm{Jamieson}, \binits{W.}},
\oauthor{\bsnm{Agrawal}, \binits{Y.}},
\oauthor{\bsnm{platipo}},
\oauthor{\bsnm{resakra}},
\oauthor{\bsnm{tal66}},
\oauthor{\bsnm{yasintoda}},
\oauthor{\bsnm{Attie}, \binits{R.}},
\oauthor{\bsnm{Murray}, \binits{S.A.}}:
2023,
SunPy,
Zenodo.
\doiurl{https://doi.org/10.5281/zenodo.7850372}.
\burl{https://doi.org/10.5281/zenodo.7850372}.
\end{botherref}
\endbibitem

\bibitem[\protect\citeauthoryear{Sarkar et~al.}{2024}]{Sarkar2024}
\begin{barticle}
\bauthor{\bsnm{Sarkar}, \binits{J.}},
\bauthor{\bsnm{Deogaonkar}, \binits{R.}},
\bauthor{\bsnm{Kesharwani}, \binits{R.}},
\bauthor{\bsnm{Padinhatteeri}, \binits{S.}},
\bauthor{\bsnm{Ramaprakash}, \binits{A.N.}},
\bauthor{\bsnm{Tripathi}, \binits{D.}},
\bauthor{\bsnm{Roy}, \binits{S.}},
\bauthor{\bsnm{Ahmed}, \binits{G.A.}},
\bauthor{\bsnm{Chatterjee}, \binits{R.}},
\bauthor{\bsnm{Ghosh}, \binits{A.}},
\bauthor{\bsnm{K.}, \binits{S.}},
\bauthor{\bsnm{Khan}, \binits{A.}},
\bauthor{\bsnm{Mehandiratta}, \binits{N.}},
\bauthor{\bsnm{Pillai}, \binits{N.}},
\bauthor{\bsnm{Singh}, \binits{S.}}:
\byear{2024},
\batitle{Science filter characterization of the Solar Ultraviolet Imaging Telescope (SUIT) on board Aditya-L1.}
\bjtitle{Experimental Astronomy}
\bvolume{59},
\bfpage{3}.
\doiurl{https://doi.org/10.1007/s10686-024-09973-5}.
\burl{https://doi.org/10.1007/s10686-024-09973-5}.
\end{barticle}
\endbibitem

\bibitem[\protect\citeauthoryear{{Sarkar et al.}}{2025}]{jj_photo}
\begin{botherref}
\oauthor{\bsnm{{Sarkar et al.}}}:
2025,
{Photometric Calbration}.
\textit{JATIS}
\textbf{submitted}.
\end{botherref}
\endbibitem

\bibitem[\protect\citeauthoryear{{Seetha} and {Megala}}{2017}]{Seetha}
\begin{barticle}
\bauthor{\bsnm{{Seetha}}, \binits{S.}},
\bauthor{\bsnm{{Megala}}, \binits{S.}}:
\byear{2017},
\batitle{{Aditya-L1 mission}}.
\bjtitle{Current Science}
\bvolume{113},
\bfpage{610}.
\doiurl{https://doi.org/10.18520/cs/v113/i04/610-612}.
\adsurl{2017CSci..113..610S}.
\end{barticle}
\endbibitem

\bibitem[\protect\citeauthoryear{{Sreejith et al.}}{2025}]{thermal2}
\begin{botherref}
\oauthor{\bsnm{{Sreejith et al.}}}:
2025,
{Thermal Filter Paper}.
\textit{\solphys}
\textbf{in prepatation}.
\end{botherref}
\endbibitem

\bibitem[\protect\citeauthoryear{{The SunPy Community} et~al.}{2020}]{sunpy20}
\begin{barticle}
\bauthor{\bsnm{{The SunPy Community}}},
\bauthor{\bsnm{Barnes}, \binits{W.T.}},
\bauthor{\bsnm{Bobra}, \binits{M.G.}},
\bauthor{\bsnm{Christe}, \binits{S.D.}},
\bauthor{\bsnm{Freij}, \binits{N.}},
\bauthor{\bsnm{Hayes}, \binits{L.A.}},
\bauthor{\bsnm{Ireland}, \binits{J.}},
\bauthor{\bsnm{Mumford}, \binits{S.}},
\bauthor{\bsnm{Perez-Suarez}, \binits{D.}},
\bauthor{\bsnm{Ryan}, \binits{D.F.}},
\bauthor{\bsnm{Shih}, \binits{A.Y.}},
\bauthor{\bsnm{Chanda}, \binits{P.}},
\bauthor{\bsnm{Glogowski}, \binits{K.}},
\bauthor{\bsnm{Hewett}, \binits{R.}},
\bauthor{\bsnm{Hughitt}, \binits{V.K.}},
\bauthor{\bsnm{Hill}, \binits{A.}},
\bauthor{\bsnm{Hiware}, \binits{K.}},
\bauthor{\bsnm{Inglis}, \binits{A.}},
\bauthor{\bsnm{Kirk}, \binits{M.S.F.}},
\bauthor{\bsnm{Konge}, \binits{S.}},
\bauthor{\bsnm{Mason}, \binits{J.P.}},
\bauthor{\bsnm{Maloney}, \binits{S.A.}},
\bauthor{\bsnm{Murray}, \binits{S.A.}},
\bauthor{\bsnm{Panda}, \binits{A.}},
\bauthor{\bsnm{Park}, \binits{J.}},
\bauthor{\bsnm{Pereira}, \binits{T.M.D.}},
\bauthor{\bsnm{Reardon}, \binits{K.}},
\bauthor{\bsnm{Savage}, \binits{S.}},
\bauthor{\bsnm{Sipőcz}, \binits{B.M.}},
\bauthor{\bsnm{Stansby}, \binits{D.}},
\bauthor{\bsnm{Jain}, \binits{Y.}},
\bauthor{\bsnm{Taylor}, \binits{G.}},
\bauthor{\bsnm{Yadav}, \binits{T.}},
\bauthor{\bsnm{Rajul}},
\bauthor{\bsnm{Dang}, \binits{T.K.}}:
\byear{2020},
\batitle{The SunPy Project: Open Source Development and Status of the Version 1.0 Core Package}.
\bjtitle{The Astrophysical Journal}
\bvolume{890}.
\doiurl{https://doi.org/10.3847/1538-4357/ab4f7a}.
\burl{https://iopscience.iop.org/article/10.3847/1538-4357/ab4f7a}.
\end{barticle}
\endbibitem

\bibitem[\protect\citeauthoryear{{Thuillier} et~al.}{2009}]{thuillier09}
\begin{barticle}
\bauthor{\bsnm{{Thuillier}}, \binits{G.}},
\bauthor{\bsnm{{Foujols}}, \binits{T.}},
\bauthor{\bsnm{{Bols{\'e}e}}, \binits{D.}},
\bauthor{\bsnm{{Gillotay}}, \binits{D.}},
\bauthor{\bsnm{{Hers{\'e}}}, \binits{M.}},
\bauthor{\bsnm{{Peetermans}}, \binits{W.}},
\bauthor{\bsnm{{Decuyper}}, \binits{W.}},
\bauthor{\bsnm{{Mandel}}, \binits{H.}},
\bauthor{\bsnm{{Sperfeld}}, \binits{P.}},
\bauthor{\bsnm{{Pape}}, \binits{S.}},
\bauthor{\bsnm{{Taubert}}, \binits{D.R.}},
\bauthor{\bsnm{{Hartmann}}, \binits{J.}}:
\byear{2009},
\batitle{{SOLAR/SOLSPEC: Scientific Objectives, Instrument Performance and Its Absolute Calibration Using a Blackbody as Primary Standard Source}}.
\bjtitle{\solphys}
\bvolume{257},
\bfpage{185}.
\doiurl{https://doi.org/10.1007/s11207-009-9361-6}.
\adsurl{2009SoPh..257..185T}.
\end{barticle}
\endbibitem

\bibitem[\protect\citeauthoryear{{Tripathi} et~al.}{2017}]{suit_main_2}
\begin{barticle}
\bauthor{\bsnm{{Tripathi}}, \binits{D.}},
\bauthor{\bsnm{{Ramaprakash}}, \binits{A.N.}},
\bauthor{\bsnm{{Khan}}, \binits{A.}},
\bauthor{\bsnm{{Ghosh}}, \binits{A.}},
\bauthor{\bsnm{{Chatterjee}}, \binits{S.}},
\bauthor{\bsnm{{Banerjee}}, \binits{D.}},
\bauthor{\bsnm{{Chordia}}, \binits{P.}},
\bauthor{\bsnm{{Gandorfer}}, \binits{A.}},
\bauthor{\bsnm{{Krivova}}, \binits{N.}},
\bauthor{\bsnm{{Nandy}}, \binits{D.}},
\bauthor{\bsnm{{Rajarshi}}, \binits{C.}},
\bauthor{\bsnm{{Solanki}}, \binits{S.K.}}:
\byear{2017},
\batitle{{The Solar Ultraviolet Imaging Telescope on-board Aditya-L1}}.
\bjtitle{Current Science}
\bvolume{113},
\bfpage{616}.
\doiurl{https://doi.org/10.18520/cs/v113/i04/616-619}.
\adsurl{2017CSci..113..616T}.
\end{barticle}
\endbibitem

\bibitem[\protect\citeauthoryear{{Tripathi} et~al.}{2023}]{aditya_mission}
\begin{bchapter}
\bauthor{\bsnm{{Tripathi}}, \binits{D.}},
\bauthor{\bsnm{{Chakrabarty}}, \binits{D.}},
\bauthor{\bsnm{{Nandi}}, \binits{A.}},
\bauthor{\bsnm{{Raghvendra Prasad}}, \binits{B.}},
\bauthor{\bsnm{{Ramaprakash}}, \binits{A.N.}},
\bauthor{\bsnm{{Shaji}}, \binits{N.}},
\bauthor{\bsnm{{Sankarasubramanian}}, \binits{K.}},
\bauthor{\bsnm{{Satheesh Thampi}}, \binits{R.}},
\bauthor{\bsnm{{Yadav}}, \binits{V.K.}}:
\byear{2023},
\bctitle{{The Aditya-L1 mission of ISRO}}.
In: \beditor{\bsnm{{Cauzzi}}, \binits{G.}},
\beditor{\bsnm{{Tritschler}}, \binits{A.}} (eds.)
\bbtitle{The Era of Multi-Messenger Solar Physics},
\bsertitle{IAU Symposium}
\bseriesno{372},
\bfpage{17}.
\doiurl{https://doi.org/10.1017/S1743921323001230}.
\adsurl{2023IAUS..372...17T}.
\end{bchapter}
\endbibitem

\bibitem[\protect\citeauthoryear{{Tripathi} et~al.}{2025}]{suit_main}
\begin{botherref}
\oauthor{\bsnm{{Tripathi}}, \binits{D.}},
\oauthor{\bsnm{{Ramaprakash}}, \binits{A.N.}},
\oauthor{\bsnm{{Padinhatteeri}}, \binits{S.}},
\oauthor{\bsnm{{Sarkar}}, \binits{J.}},
\oauthor{\bsnm{{Burse}}, \binits{M.}},
\oauthor{\bsnm{{Tyagi}}, \binits{A.}},
\oauthor{\bsnm{{Kesharwani}}, \binits{R.}},
\oauthor{\bsnm{{Sinha}}, \binits{S.}},
\oauthor{\bsnm{{Joshi}}, \binits{B.}},
\oauthor{\bsnm{{Deogaonkar}}, \binits{R.}},
\oauthor{\bsnm{{Roy}}, \binits{S.}},
\oauthor{\bsnm{{Nived}}, \binits{V.N.}},
\oauthor{\bsnm{{Gopalakrishnan}}, \binits{R.}},
\oauthor{\bsnm{{Kulkarni}}, \binits{A.}},
\oauthor{\bsnm{{Khan}}, \binits{A.}},
\oauthor{\bsnm{{Ghosh}}, \binits{A.}},
\oauthor{\bsnm{{Rajarshi}}, \binits{C.}},
\oauthor{\bsnm{{Modi}}, \binits{D.}},
\oauthor{\bsnm{{Kumar}}, \binits{G.}},
\oauthor{\bsnm{{Yadav}}, \binits{R.}},
\oauthor{\bsnm{{Varma}}, \binits{M.}},
\oauthor{\bsnm{{Bayanna}}, \binits{R.}},
\oauthor{\bsnm{{Chordia}}, \binits{P.}},
\oauthor{\bsnm{{Karmakar}}, \binits{M.}},
\oauthor{\bsnm{{Abraham}}, \binits{L.}},
\oauthor{\bsnm{{Adithya}}, \binits{H.N.}},
\oauthor{\bsnm{{Adoni}}, \binits{A.}},
\oauthor{\bsnm{{Ahmed}}, \binits{G.A.}},
\oauthor{\bsnm{{Banerjee}}, \binits{D.}},
\oauthor{\bsnm{{Ram}}, \binits{B.}},
\oauthor{\bsnm{{Bhandare}}, \binits{R.}},
\oauthor{\bsnm{{Chatterjee}}, \binits{S.}},
\oauthor{\bsnm{{Chillal}}, \binits{K.}},
\oauthor{\bsnm{{Dey}}, \binits{A.}},
\oauthor{\bsnm{{Gandorfer}}, \binits{A.}},
\oauthor{\bsnm{{Gowda}}, \binits{G.}},
\oauthor{\bsnm{{Haridas}}, \binits{T.R.}},
\oauthor{\bsnm{{Jain}}, \binits{A.}},
\oauthor{\bsnm{{James}}, \binits{M.}},
\oauthor{\bsnm{{Jayakumar}}, \binits{R.P.}},
\oauthor{\bsnm{{Leeja Justin}}, \binits{E.}},
\oauthor{\bsnm{{Nagaraju}}, \binits{K.}},
\oauthor{\bsnm{{Kathait}}, \binits{D.}},
\oauthor{\bsnm{{Khodade}}, \binits{P.}},
\oauthor{\bsnm{{Kiran}}, \binits{M.}},
\oauthor{\bsnm{{Kohok}}, \binits{A.}},
\oauthor{\bsnm{{Krivova}}, \binits{N.}},
\oauthor{\bsnm{{Kumar}}, \binits{N.}},
\oauthor{\bsnm{{Mehandiratta}}, \binits{N.}},
\oauthor{\bsnm{{Mestry}}, \binits{V.}},
\oauthor{\bsnm{{Motamarri}}, \binits{S.}},
\oauthor{\bsnm{{Mustafa}}, \binits{S.F.}},
\oauthor{\bsnm{{Nandy}}, \binits{D.}},
\oauthor{\bsnm{{Narendra}}, \binits{S.}},
\oauthor{\bsnm{{Navle}}, \binits{S.}},
\oauthor{\bsnm{{Parate}}, \binits{N.}},
\oauthor{\bsnm{{Pillai}}, \binits{A.M.}},
\oauthor{\bsnm{{Punnadi}}, \binits{S.}},
\oauthor{\bsnm{{Rajendra}}, \binits{A.}},
\oauthor{\bsnm{{Ravi}}, \binits{A.}},
\oauthor{\bsnm{{Raha}}, \binits{B.}},
\oauthor{\bsnm{{Sankarasubramanian}}, \binits{K.}},
\oauthor{\bsnm{{Sarvar}}, \binits{G.}},
\oauthor{\bsnm{{Shaji}}, \binits{N.}},
\oauthor{\bsnm{{Sharma}}, \binits{N.}},
\oauthor{\bsnm{{Singh}}, \binits{A.}},
\oauthor{\bsnm{{Singh}}, \binits{S.}},
\oauthor{\bsnm{{Solanki}}, \binits{S.K.}},
\oauthor{\bsnm{{Subramanian}}, \binits{V.}},
\oauthor{\bsnm{{T}}, \binits{R.}},
\oauthor{\bsnm{{T}}, \binits{S.}},
\oauthor{\bsnm{{Thatimattala}}, \binits{S.}},
\oauthor{\bsnm{{Krishna Tota}}, \binits{H.}},
\oauthor{\bsnm{{TS}}, \binits{V.}},
\oauthor{\bsnm{{Unnikrishnan}}, \binits{A.}},
\oauthor{\bsnm{{Vadodariya}}, \binits{K.}},
\oauthor{\bsnm{{Veeresha}}, \binits{D.R.}},
\oauthor{\bsnm{{Venkateswaran}}, \binits{R.}}:
2025,
{The Solar Ultraviolet Imaging Telescope on board Aditya-L1}.
\textit{arXiv e-prints},
arXiv:2501.02274.
\doiurl{https://doi.org/10.48550/arXiv.2501.02274}.
\adsurl{2025arXiv250102274T}.
\end{botherref}
\endbibitem

\bibitem[\protect\citeauthoryear{{Varma} et~al.}{2023}]{suit_algo}
\begin{barticle}
\bauthor{\bsnm{{Varma}}, \binits{M.}},
\bauthor{\bsnm{{u'Padinhatteeri}}, \binits{S.}},
\bauthor{\bsnm{{Sinha}}, \binits{S.}},
\bauthor{\bsnm{{Tyagi}}, \binits{A.}},
\bauthor{\bsnm{{Burse}}, \binits{M.}},
\bauthor{\bsnm{{Yadav}}, \binits{R.}},
\bauthor{\bsnm{{Kumar}}, \binits{G.}},
\bauthor{\bsnm{{Ramaprakash}}, \binits{A.}},
\bauthor{\bsnm{{Tripathi}}, \binits{D.}},
\bauthor{\bsnm{{Sankarasubramanian}}, \binits{K.}},
\bauthor{\bsnm{{Nagaraju}}, \binits{K.}},
\bauthor{\bsnm{{Vadodariya}}, \binits{K.}},
\bauthor{\bsnm{{Tadepalli}}, \binits{S.}},
\bauthor{\bsnm{{Deogaonkar}}, \binits{R.}},
\bauthor{\bsnm{{Olekar}}, \binits{M.}},
\bauthor{\bsnm{{Azaruddin}}, \binits{M.}},
\bauthor{\bsnm{{Unnikrishnan}}, \binits{A.}}:
\byear{2023},
\batitle{{The Solar Ultra-Violet Imaging Telescope (SUIT) Onboard Intelligence for Flare Observations}}.
\bjtitle{\solphys}
\bvolume{298},
\bfpage{16}.
\doiurl{https://doi.org/10.1007/s11207-023-02108-7}.
\adsurl{2023SoPh..298...16V}.
\end{barticle}
\endbibitem

\bibitem[\protect\citeauthoryear{Virtanen et~al.}{2020}]{scipy}
\begin{barticle}
\bauthor{\bsnm{Virtanen}, \binits{P.}},
\bauthor{\bsnm{Gommers}, \binits{R.}},
\bauthor{\bsnm{Oliphant}, \binits{T.E.}},
\bauthor{\bsnm{Haberland}, \binits{M.}},
\bauthor{\bsnm{Reddy}, \binits{T.}},
\bauthor{\bsnm{Cournapeau}, \binits{D.}},
\bauthor{\bsnm{Burovski}, \binits{E.}},
\bauthor{\bsnm{Peterson}, \binits{P.}},
\bauthor{\bsnm{Weckesser}, \binits{W.}},
\bauthor{\bsnm{Bright}, \binits{J.}},
\bauthor{\bsnm{{van der Walt}}, \binits{S.J.}},
\bauthor{\bsnm{Brett}, \binits{M.}},
\bauthor{\bsnm{Wilson}, \binits{J.}},
\bauthor{\bsnm{Millman}, \binits{K.J.}},
\bauthor{\bsnm{Mayorov}, \binits{N.}},
\bauthor{\bsnm{Nelson}, \binits{A.R.J.}},
\bauthor{\bsnm{Jones}, \binits{E.}},
\bauthor{\bsnm{Kern}, \binits{R.}},
\bauthor{\bsnm{Larson}, \binits{E.}},
\bauthor{\bsnm{Carey}, \binits{C.J.}},
\bauthor{\bsnm{Polat}, \binits{{\. I}.}},
\bauthor{\bsnm{Feng}, \binits{Y.}},
\bauthor{\bsnm{Moore}, \binits{E.W.}},
\bauthor{\bsnm{{VanderPlas}}, \binits{J.}},
\bauthor{\bsnm{Laxalde}, \binits{D.}},
\bauthor{\bsnm{Perktold}, \binits{J.}},
\bauthor{\bsnm{Cimrman}, \binits{R.}},
\bauthor{\bsnm{Henriksen}, \binits{I.}},
\bauthor{\bsnm{Quintero}, \binits{E.A.}},
\bauthor{\bsnm{Harris}, \binits{C.R.}},
\bauthor{\bsnm{Archibald}, \binits{A.M.}},
\bauthor{\bsnm{Ribeiro}, \binits{A.H.}},
\bauthor{\bsnm{Pedregosa}, \binits{F.}},
\bauthor{\bsnm{{van Mulbregt}}, \binits{P.}},
\bauthor{\bsnm{{SciPy 1.0 Contributors}}}:
\byear{2020},
\batitle{{{SciPy} 1.0: Fundamental Algorithms for Scientific Computing in Python}}.
\bjtitle{Nature Methods}
\bvolume{17},
\bfpage{261}.
\doiurl{https://doi.org/10.1038/s41592-019-0686-2}.
\adsurl{https://rdcu.be/b08Wh}.
\end{barticle}
\endbibitem

\end{thebibliography}
\end{document}